\documentclass[twocolumn]{aastex6}

\usepackage{color}

\begin{document}

\title{Planetesimal formation by the streaming instability in a photoevaporating disk}

\author{Daniel Carrera\altaffilmark{1}, Uma Gorti\altaffilmark{2,3}, Anders Johansen\altaffilmark{1}, and Melvyn B. Davies\altaffilmark{1}}

\noaffiliation

\altaffiltext{1}{Lund Observatory, Dept of Astronomy and Theoretical Physics, Lund University, Box 43, SE-221 00 Lund, Sweden}
\altaffiltext{2}{NASA Ames Research Center, Moffett Field, CA, USA}
\altaffiltext{3}{SETI Institute, Mountain View, CA, USA}

%
% New commands
%
\newcommand{\K}{\, {\rm K}}
\newcommand{\at}{\, \alpha_{\rm t}}
\newcommand{\av}{\, \alpha_{\rm v}}
\newcommand{\uf}{\, u_{\rm frag}}
\newcommand{\ms}{\, {\rm m \, s}^{-1}}
\newcommand{\cm}{\, {\rm cm}}
\newcommand{\St}{\, {\rm St}}
\newcommand{\kb}{\, {\rm k_B}}
\newcommand{\dyne}{\, {\rm dyne}}

\newcommand{\pe}{{\rm pe}}
\newcommand{\erf}{{\rm erf}}

\newcommand{\rg}{\, r_{\rm g}}
\newcommand{\rs}{\, r_{\rm s}}
\newcommand{\ns}{\, n_{\rm s}}
\newcommand{\cs}{\, c_{\rm s}}
\newcommand{\ts}{\, t_{\rm s}}
\newcommand{\cms}{\, {\rm cm \, s}^{-1}}
\newcommand{\kms}{\, {\rm km \, s}^{-1}}
\newcommand{\rhos}{\, \rho_{\rm s}}
\newcommand{\rhog}{\, \rho_{\rm g}}
\newcommand{\rhop}{\, \rho_{\rm p}}
\newcommand{\rhoR}{\, \rho_{\rm R}}
\newcommand{\ergs}{\, {\rm erg} \; {\rm s}^{-1}}
\newcommand{\Zcrit}{\, Z_{\rm crit}}
\newcommand{\Omegak}{\, \Omega_{\rm K}}
\newcommand{\Sigmag}{\, \Sigma_{\rm g}}
\newcommand{\Sigmap}{\, \Sigma_{\rm p}}
\newcommand{\Sigmad}{\, \Sigma_{\rm d}}
\newcommand{\md}{\, m_{\rm d}}

\newcommand{\blue}[1]{{\color{blue} #1}}

%% Mark off the abstract in the ``abstract'' environment. 
\begin{abstract}
%
% 1) Context.
%
Recent years have seen growing interest in the streaming instability as a candidate mechanism to produce planetesimals. However, these investigations have been limited to small-scale simulations.
%
% 3) What did I do?
%
We now present the results of a global protoplanetary disk evolution model that incorporates planetesimal formation by the streaming instability, along with viscous accretion, photoevaporation by EUV, FUV, and X-ray photons, dust evolution, the water ice line, and stratified turbulence.
%
% 4) What did I find?
%
Our simulations produce massive (60-130 $M_\oplus$) planetesimal belts beyond 100 au and up to $\sim 20 M_\oplus$ of planetesimals in the middle regions (3-100 au). Our most comprehensive model forms 8 $M_\oplus$ of planetesimals inside 3 au, where they can give rise to terrestrial planets. The planetesimal mass formed in the inner disk depends critically on the timing of the formation of an inner cavity in the disk by high-energy photons.
%
% 5) What does it mean?
%
Our results show that the combination of photoevaporation and the streaming instability are efficient at converting the solid component of protoplanetary disks into planetesimals.
%
% 6) What is the next step?
%
Our model, however, does not form enough early planetesimals in the inner and middle regions of the disk to give rise to giant planets and super-Earths with gaseous envelopes. Additional processes such as particle pileups and mass loss driven by MHD winds may be needed to drive the formation of early planetesimal generations in the planet forming regions of protoplanetary disks.
\end{abstract}


%%%%%%%%%%%%%%%%%%%%%%%%%%%%%%%%%%%%%%%%%%%%%%%%%%%%%%%%%%%%%%%%%%%%%%%%%%%%%%%
%
% INTRODUCTION
%
%%%%%%%%%%%%%%%%%%%%%%%%%%%%%%%%%%%%%%%%%%%%%%%%%%%%%%%%%%%%%%%%%%%%%%%%%%%%%%%
\section{Introduction}
\label{sec:intro}
Most young stars are surrounded by a circumstellar disk of gas and dust with a mass between 0.01\% and 10\% that of the central star \citep{Andrews_2005}. The disk has a characteristic lifetime of 2-5 Myr, with a significant scatter due to variations in stellar mass and local environment  \citep[e.g.][]{Hartmann_1998,Haisch_2001,Mamajek_2009}. Thanks to radial velocity surveys and the results of the Kepler mission it is now established that planets between the mass of Earth and Neptune (``super-Earths'') are among the most abundant in the Galaxy \citep{Howard_2010,Borucki_2011,Batalha_2013}. From the bulk density of these planets we know that many of them must possess large H/He envelopes \citep[e.g.][]{Adams_2008,Lopez_2014} which must have been accreted onto a solid core before the disk dissipated.  In the pebble accretion model, the solid core is produced by the accretion of cm-size ``pebbles'' onto a seed planetesimal \citep[e.g.][]{Ormel_2010, Lambrechts_2012, Bitsch_2015, Levison_2015}. After the disk dissipates, the region close to the star is thought to be populated by planetesimals and Mars-sized planetary embryos \citep{Kokubo_1996,Johansen_2015}. These bodies are the seed material for the formation of terrestrial planets over the next 100 Myr \citep[e.g.][]{Chambers_1998,Raymond_2006}. This has two implications for planetesimals:

\begin{itemize}
\item Some planetesimals should form early to allow sufficient time for the formation of super-Earths and giant planet cores before the disk dissipates. These cores experience disk migration, so they must have formed at greater distances than where we see them today. For example, a Jupiter-mass planet at 3 au may come from a seed that formed in the $20-30$ au region \citep{Bitsch_2015}.

\item Some planetesimals must form dry --- inside the water ice line --- in order to reproduce the very low water content of the terrestrial planets in the solar system. The water present in the terrestrial planets likely comes from a late accretion of icy planetesimals scattered inward \citep{Raymond_2004}.
\end{itemize}

In other words, a complete model of planetesimal formation should be able to produce planetesimals beyond the water ice line early, and it must also be able to produce dry planetesimals in the terrestrial zone at some point (possibly late). Furthermore, exoplanet observations suggest a wide range of super-Earth compositions, which may indicate a diversity in their formation pathways and perhaps also times \citep[e.g.][]{Lopez_2014}.

The aim of this paper is to address the formation of planetesimals in a protoplanetary disk model that includes a state-of-the-art description of photoevaporation \citep[based on the recent model of][]{Gorti_2015}. Photoevaporation may be key to forming planetesimals by the streaming instability, because pebble-sized particles can only concentrate in regions of elevated metallicity compared to the nominal 1\% value of the solar photosphere \citep{Johansen_2009,Bai_2010,Carrera_2015}. In a complementary work, \citet{Drazkowska_2016} also modelled the evolution of dust in an evolving disk. The most salient difference between our work and theirs is that we use a more sophisticated model for photoevaporation. \citet{Drazkowska_2016} assume a fixed mass-loss rate that is insensitive to changes in accretion, or the radial and size distribution of dust. Our model is described in sections \ref{sec:model:photoevap} and \ref{sec:methods:NP}. In a real protoplanetary disk, the FUV flux is dominated by photons released in the accretion process, and the resulting FUV heating depends in a complicated way on grain size and distribution.

Another important difference between our work and \citet{Drazkowska_2016} is that we use the turbulent viscosity motivated by observation. In a turbulent disk with viscosity $\nu = \av \cs H$ \citep{Shakura_1976} where $\cs$ is the local sound speed and $\av$ is the dimension-less turbulence parameter, the timescale for viscous evolution is

\begin{equation}\label{eqn:t_nu}
	t_\nu = \frac{1}{\av \; \Omega} \left( \frac{H}{R} \right)^{-2},
\end{equation}
where $\Omega$ is the Keplerian frequency and $H/R \sim 0.1$ is the disk aspect ratio \citep[e.g.][]{Alexander_2014}. For disk lifetimes of a few Myr, we obtain $\av \sim 10^{-2}$, and that is the value we use \citep[see also][]{Hartmann_1998,Alexander_2014}. In contrast, \citet{Drazkowska_2016} set $\av$ to $10^{-3}$ to facilitate planetesimal formation. We also allow smaller particles to participate in the streaming instability, following the results of \citet{Carrera_2015}; however this is a minor difference. One interesting feature of \citet{Drazkowska_2016} is that the radial drift of solids decreases with increasing dust-to-gas ratio, as they include the back-reaction of the dust on the gas in their models. This facilitates the pile-up of solids in the inner disk, and may play a role in planetesimal formation. In an upcoming paper we will combine our more sophisticated models for photoevaporation and streaming instability with the more sophisticated radial drift model of \citet{Drazkowska_2016}.

This paper is organized as follows. In section \ref{sec:model} we introduce main concepts pertaining to our disk evolution model. In section \ref{sec:methods} we explain how the model is implemented and our key assumptions. Our results are presented in section \ref{sec:results}, where we show that planetesimal formation is a natural by-product of the evolution of a protoplanetary disk. We show that planetesimal formation in the outer disk is a natural outcome of disk evolution. We also show that standard prescriptions for disk processes can lead to several Earth masses of planetesimals forming in the terrestrial zone. Then in section \ref{sec:other} we discuss various processes that we could not model adequately in our simulations, and speculate on how they might affect our results. Finally, we summarize our results and draw conclusions in section \ref{sec:conclusions}.

%%%%%%%%%%%%%%%%%%%%%%%%%%%%%%%%%%%%%%%%%%%%%%%%%%%%%%%%%%%%%%%%%%%%%%%%%%%%%%%
%
% MODEL DESCRIPTION
%
%%%%%%%%%%%%%%%%%%%%%%%%%%%%%%%%%%%%%%%%%%%%%%%%%%%%%%%%%%%%%%%%%%%%%%%%%%%%%%%
\section{Main Theoretical Concepts}
\label{sec:model}

In the interstellar medium most of the solids are in the form of sub-micron dust \citep[e.g.][]{Weingartner_2001,Lefevre_2014}. Inside the protoplanetary disk, these grains stick and grow to larger sizes up to at least millimetre sizes in the outer disk \citep{van_der_Marel_2013}. Growth beyond a few millimetres is inhibited by  collisional fragmentation \citep{Brauer_2008}. The mechanism by which these sub-cm grains are converted into 100-km planetesimals is not well understood.

%-----------------------------------------------------------------------------%
% RADIAL DRIFT
%-----------------------------------------------------------------------------%
\subsection{The streaming instability}
\label{sec:model:SI}

Many current models seek to produce an over-density of dust grains dense enough to self-gravitate and undergo gravitational collapse \citep[e.g.][]{Safronov_1969,Goldreich_1973,Johansen_2014}. These over-densities may be produced by various hydrodynamic processes including long-lived gaseous vortices \citep{Barge_1995,Meheut_2012}, pressure bumps \citep{Johansen_2009,Johansen_2011,Simon_2012}, or through the run-away accumulation of particles by radial drift, known as the streaming instability \citep[e.g.][]{Youdin_2005,Johansen_2007,Bai_2010}. The streaming instability has drawn recent attention because it has proven effective at reaching very high particle densities and forming planetesimals on very short (orbital) time scales. In this work, we consider the formation of planetesimals via the streaming instability.

The way that a solid grain responds to aerodynamic drag is determined by its Stokes number, $\St = \ts \Omegak$, where $\ts$ is the stopping time and $\Omegak$ is the Keplerian frequency. For particles in the Epstein regime (particles smaller than the mean free path of gas) the stopping time is given by,

\begin{equation}\label{eqn:St}
	\ts = \frac{\rhos a}{\rho \cs} \sqrt{\frac{\pi}{8}},
\end{equation}
where $\rhos$ is the material density, $\rho$ is the local gas density, $a$ is the grain size, and $\cs$ is the local sound speed (we have verified that the particles in our simulations are always in the Epstein regime.) \citet{Carrera_2015} have shown that the streaming instability can be effective at producing particle clumps for a wide range of Stokes numbers, but grains smaller than $\St = 0.1$ require increasingly large densities for the streaming instability to occur. Their measurement of the threshold metallicity as a function of the Stokes number will be used in this work to incorporate planetesimal formation into a global model of an evolving protoplanetary disk. Our prescription for forming planetesimals will be discussed further in section \ref{sec:methods:SI}.

%-----------------------------------------------------------------------------%
% COAGULATION AND FRAGMENTATION -- removed \subsection{} and \label{}
%-----------------------------------------------------------------------------%

%-----------------------------------------------------------------------------%
% DUST GROWTH
%-----------------------------------------------------------------------------%
\subsection{Dust growth}
\label{sec:model:dust}

The size of the grains is determined by the balance between dust coagulation, collisional fragmentation, and radial drift \citep[e.g.][]{Birnstiel_2012}. Silicate grains also experience a bouncing barrier \citep{Zsom_2010}. In a standard $\alpha$-disk \citep{Shakura_1976}, turbulence generates viscosity which is parametrized by a parameter $\av$. In the simplest models, turbulence is assumed to be uniform and the turbulence parameter $\at = \av$ (from Equation (\ref{eqn:t_nu})).  The collision speed between small particles is

\begin{equation}\label{eqn:du_coll}
	\Delta u_{\rm coll} \sim \cs \sqrt{\at \St}.
\end{equation}
Here, we make a distinction between the midplane turbulence $\at$ and the global $\av$ --- which sets the viscous evolution of the disk --- since $\at$ and $\av$ may differ in a real disk. This formula (Equation (\ref{eqn:du_coll})) is valid for St roughly between $10^{-3}$ and $1$ \citep[e.g.][]{Ormel_2008,Weidenschilling_1984}. This means that, in the fragmentation-limited regime, the Stokes number of the largest particles is independent of the gas density,

\begin{equation}\label{eqn:Stfrag}
	\St_{\rm frag} \sim \frac{\uf^2}{\at \cs^2},
\end{equation}
where $\uf$ is the fragmentation speed \citep{Birnstiel_2009}. In other words, as the disk evolves and the gas density drops, any increase in Stokes number (Equation (\ref{eqn:St})) immediately leads to greater collision speeds (Equation (\ref{eqn:du_coll})) so that the particles fragment until the Stokes number once again reaches $\St_{\rm frag}$ (Equation (\ref{eqn:Stfrag})). As a canonical example, $\uf = 10 \ms$, $\cs = 800 \ms$, and $\av = \at = 10^{-2}$ would give $\St_{\rm frag} \sim 0.01$, independent of the gas density. The $\uf = 10 \ms$ value is consistent with ice-dust aggregates \citep{Wada_2009,Gundlach_2015}. If most of the mass in the dust is in grains close to this maximum size, the streaming instability requires a dust-to-gas ratio of at least $\Sigmap / \Sigmag \sim 0.03$ \citep{Carrera_2015}.

Given that the dust-to-gas ratio in the ISM is closer to $\sim 0.01$ \citep[e.g.][]{Leroy_2013}, the streaming instability requires processes that either concentrates dust locally, or that preferentially removes the gas component of the disk while leaving the dust behind. We now discuss some mechanisms that can accomplish the required dust concentration.

%-----------------------------------------------------------------------------%
% RADIAL DRIFT
%-----------------------------------------------------------------------------%
\subsection{Radial drift and the water ice line}
\label{sec:model:drift}

Inside a protoplanetary disk, solid particles experience a head wind as the gas component orbits at a sub-Keplerian speed. The aerodynamic drag leads to a loss of angular momentum, causing the solids to drift toward the star. The drift velocity is given by

\begin{equation}\label{eqn:u_drift}
	u_{\rm drift} = \frac{-2 \Delta v}{\St + \St^{-1}}	
\end{equation}
where $\Delta v = u_{\rm kep} - u_{\rm gas}$ is the head wind speed \citep{Weidenschilling_1977}. Particles that are either very small or very large are relatively unaffected by radial drift. The fastest drift speed occurs for St = 1.  In some cases, the collision speed $\Delta u_{\rm coll}$ is dominated by differential drift. The relative drift speed for small particles ($\St \ll 1$) is given by the expression

\begin{equation}\label{eqn:Delta_u_drift}
	\Delta u_{\rm drift} = \frac{\cs^2}{u_{\rm kep}}
    		\left| \frac{\partial \ln P}{\partial \ln r} \right| |\St_1 - \St_2|.
\end{equation}
Following the notation of \citet{Birnstiel_2012}, we write $\St_2 = N\St_1 < \St_1$. Setting $\Delta u_{\rm drift} = u_{\rm frag}$ one gets the drift-driven fragmentation limit,

\begin{equation}\label{eqn:St_df}
	\St_{\rm df} =  \frac{u_{\rm frag} u_{\rm kep}}{\cs^2 (1-N)}
	    		\left| \frac{\partial \ln P}{\partial \ln r} \right|^{-1},
\end{equation}
where $P$ is the gas pressure and $r$ is the radial coordinate \citep{Birnstiel_2012}. This quantity depends weakly on $r$, and $\St_{\rm df} \approx 0.19$ for most of the disk. In some cases, particles drift before they can reach the fragmentation limit. The maximum Stokes number that grains can reach in a drift-dominated region is

\begin{equation}\label{eqn:Stdrift}
	\St_{\rm drift} \sim \frac{\Sigmap}{\Sigmag} \frac{u_{\rm kep}^2}{\cs^2}
    					\left| \frac{\partial \ln P}{\partial \ln r} \right|^{-1}.
\end{equation}
In the outer disk, where $u_{\rm kep}$ is small, $\St_{\rm drift} < \St_{\rm frag}$ and dust growth is drift-limited. However, for our model parameters, fragmentation by differential drift is the more stringent condition.

The water ice line may also play a role in transferring solids into the inner disk. Beyond the ice line, solids would be in the form of ice-silicate aggregates. When they drift across the ice line, the ice component sublimates, leaving behind the silicates. If silicates have a lower fragmentation speed than ice, or if their growth is limited by a bouncing barrier \citep{Zsom_2010}, they would be unable to grow to their former size to start drifting again. This process could result in an additional accumulation of dust grains in the vicinity of the water ice line \citep{Sirono_2011b}. However, the viscous accretion speed of gas and particles can limit the degree of pile-up.

%-----------------------------------------------------------------------------%
% PHOTOEVAPORATION
%-----------------------------------------------------------------------------%
\subsection{Photoevaporation}
\label{sec:model:photoevap}

Another way to increase the dust-to-gas ratio and trigger the streaming instability is to preferentially remove the gas component while leaving the solids in the disk. A well studied mechanism that can accomplish this is photoevaporation. It occurs when energetic photons, primarily in the form of FUV, EUV, and X-rays, heat gas to the point that it becomes gravitationally unbound from the star.

EUV heating occurs through the direct ionization of hydrogen atoms. The absorption cross section at the Lyman limit is large, and drops quickly for more energetic photons. The ionized layer is heated to $T \sim 10^4 \K$ ($\cs \sim 10 \kms$), which can launch a flow at $\sim 1$ au for a star of mass 1 $M_\odot$. When the photoevporative mass loss rate exceeds the local accretion rate, a gap forms in the disk, and viscous accretion depletes the inner disk to form a hole. The mass loss rate due to EUV depends almost exclusively on the flux at $\sim 1$ au. However, the intrinsic EUV luminosity is poorly determined and the flux incident on the disk is very uncertain because the accretion column, as well as any jets or winds close to the star are extremely optically thick to EUV \citep{Alexander_2014}. Once the disk forms an optically thin cavity, direct irradiation onto the inner edge produces an order-of-magnitude increase in the photoevaporation rate and the remaining disk dissipates rapidly \citep{Alexander_2006a, Alexander_2006b}.

X-ray heating occurs indirectly, through the K-shell ionization of heavy elements (mainly O, C, and Fe), where ejected electrons heat the hydrogen gas. T Tauri stars are bright X-ray sources with a median luminosity of $L_X \sim 10^{30} \ergs$. The photon energy ranges from $\sim 0.1-10$ keV and gas temperatures from a few 100 K to over $10^4$ K. The softer X-ray photons heat the gas to higher temperatures ($T \sim 1-2\times 10^4 \K$) but are also easily absorbed at small column densities. The mass loss profile is wider than EUV, with a peak around 3 au \citep{Alexander_2014}. While the X-ray photon flux is better determined than that of EUV, the heating process is much more complex. The main uncertainties come from imprecise knowledge of the soft X-ray flux incident on the disk and assumptions about disk chemistry which affect gas cooling calculations \citep{Alexander_2014}.

FUV photoevaporation is by far the most complex. Heating mainly occurs indirectly through the emission of photo-electrons from small dust grains and polycyclic aromatic hydrocarbon (PAH) molecules, which in turn collide with hydrogen gas. Because dust grains provide the opacity, FUV heating depends not only on the FUV flux, but also on the abundance of small dust grains. The cooling further depends on gas chemistry, similar to the X-ray case, and FUV mass loss rates are subject to the same uncertainties.

Young stars have active chromospheres and hence powerful high-energy photon fluxes across the X-ray, EUV and FUV bands. In addition, accretion shocks due to flows onto the stellar surface at early epochs of star formation can generate a high FUV flux \citep[e.g.][]{Gullbring_2000,Ingebly_2011,Yang_2012}. \citet{Gorti_2009b} found that this accretion-dominated FUV flux can cause photoevaporation during very early stages of disk evolution. It is unclear whether the accretion component has substantial contributions to EUV and X-rays, and if these photons can penetrate the accretion columns and irradiate the disk \citep{Gudel_2007}. \citet{Gorti_2009b} argue that EUV and soft X-rays are strongly absorbed when the disk accretion rates are high and only become relevant at late stages of evolution, and found that X-rays can be important if they have a predominantly soft spectrum.

Mass loss due to FUV photoevaporation occurs primarily in the inner disk ($\sim 3-10$ au) and in the outer disk ($\gtrsim 50 $au). FUV photoevaporation in the outer disk is particularly significant for disk evolution as this is where most of the disk mass resides and where temperatures as low as 100 K can unbind the hydrogen gas \citep{Gorti_2015}. Thus, FUV photoevaporation truncates the outer edge of the disk even at early epochs. As the disk dissipates and accretion drops, the FUV luminosity is also mainly chromospheric and  becomes more comparable to EUV and X-rays. As noted earlier, once an optically thin cavity forms, FUV, EUV and X-rays push the cavity outward even as FUV continues to bring the outer edge inward. The two evaporation fronts meet in the middle, and eventually completely disperse the disk \citep{Gorti_2015}.

%-----------------------------------------------------------------------------%
% BIRNSTIEL MODEL AT HIGH DUST-TO-GAS RATIOS
%-----------------------------------------------------------------------------%
\subsection{Dust evolution at high dust-to-gas ratio}
\label{sec:model:caveat}

The dust evolution model that we presented in this section was derived by \citet{Birnstiel_2012} under an assumption that the gas is dynamically dominant medium. Photoevaporation could bring the dust-to-gas ratio above unity, so that the assumptions of the model no longer apply. If the dust-to-gas ratio \textit{were} to move past unity, we would expect the solid mass to dampen the gas turbulence which would lead to lower collision speeds and larger particles. However, we expect that the streaming instability prevents the dust-to-gas ratio from ever reaching unity in the first place. We will see in section \ref{sec:methods:SI} that the streaming instability has a critical dust-to-gas ratio that is less than unity, and above that value any excess dust efficiently converted into planetesimals.

%-----------------------------------------------------------------------------%
% DISK WINDS
%-----------------------------------------------------------------------------%
\subsection{Disk winds}
\label{sec:model:winds}

Although we do not include disk winds in our base model, we will discuss them briefly here. Disk winds are are driven by a combination of thermal heating and magnetocentrifugal acceleration along large-scale magnetic fields going through the disk. As such, these winds exert a torque on the disk so they carry a significant portion of the disk's angular momentum and drive much of the accretion. The process was first proposed by \citet{Blandford_1982}, but see also the review by \citet{Turner_2014} and the recent model by \citet{Bai_2016a}.

Similar to photoevaporation, disk winds should preferentially remove gas rather than solids. \citet{Bai_2016b} has noted that mass loss predominantly occurs in the outer disk and that the corresponding increase in $\Sigmap/\Sigmag$ may promote planetesimal formation by the streaming instability.

%%%%%%%%%%%%%%%%%%%%%%%%%%%%%%%%%%%%%%%%%%%%%%%%%%%%%%%%%%%%%%%%%%%%%%%%%%%%%%%
%
% METHODS
%
%%%%%%%%%%%%%%%%%%%%%%%%%%%%%%%%%%%%%%%%%%%%%%%%%%%%%%%%%%%%%%%%%%%%%%%%%%%%%%%
\section{Methods}
\label{sec:methods}

All our simulations begin at the end of the Class I phase, after about 0.5 to 1 Myr of evolution. At this stage the central star has acquired most of its final mass and the envelope has dissipated to reveal a gravitationally stable, accreting, protoplanetary disk \citep{Bell_2013,Dunham_2014}. The model parameters are all as described in \citet{Gorti_2015}, as in their fiducial run. We briefly summarize some of these here for completion. The initial disk mass is $0.1 M_\odot$, distributed in a $1/r$ power law extending to 200au; this profile is gravitationally stable everywhere. The disk is allowed to viscously relax before the simulations are begun, and attains a self-similar (viscous) radial profile (in $\sim 10^5$ years). The gas-to-dust ratio is initially 100, and the dust sizes (32 bins, in the range 50\AA-1cm) are computed as the disk evolves. The 1-D disk evolution model is solved using a Crank-Nicholson method with an additional criterion imposed by the rate of change of surface density due to photoevaporation ($\Delta t < 0.1\Sigma/\dot{\Sigma}$). The temperature and density structure of the disk is determined from the instantaneous surface density profile ($\Sigma(r,t)$) using a 1+1D thermochemical disk model \citep{Gorti_2015}.

\begin{figure}[h!]
	\includegraphics[width=\columnwidth]{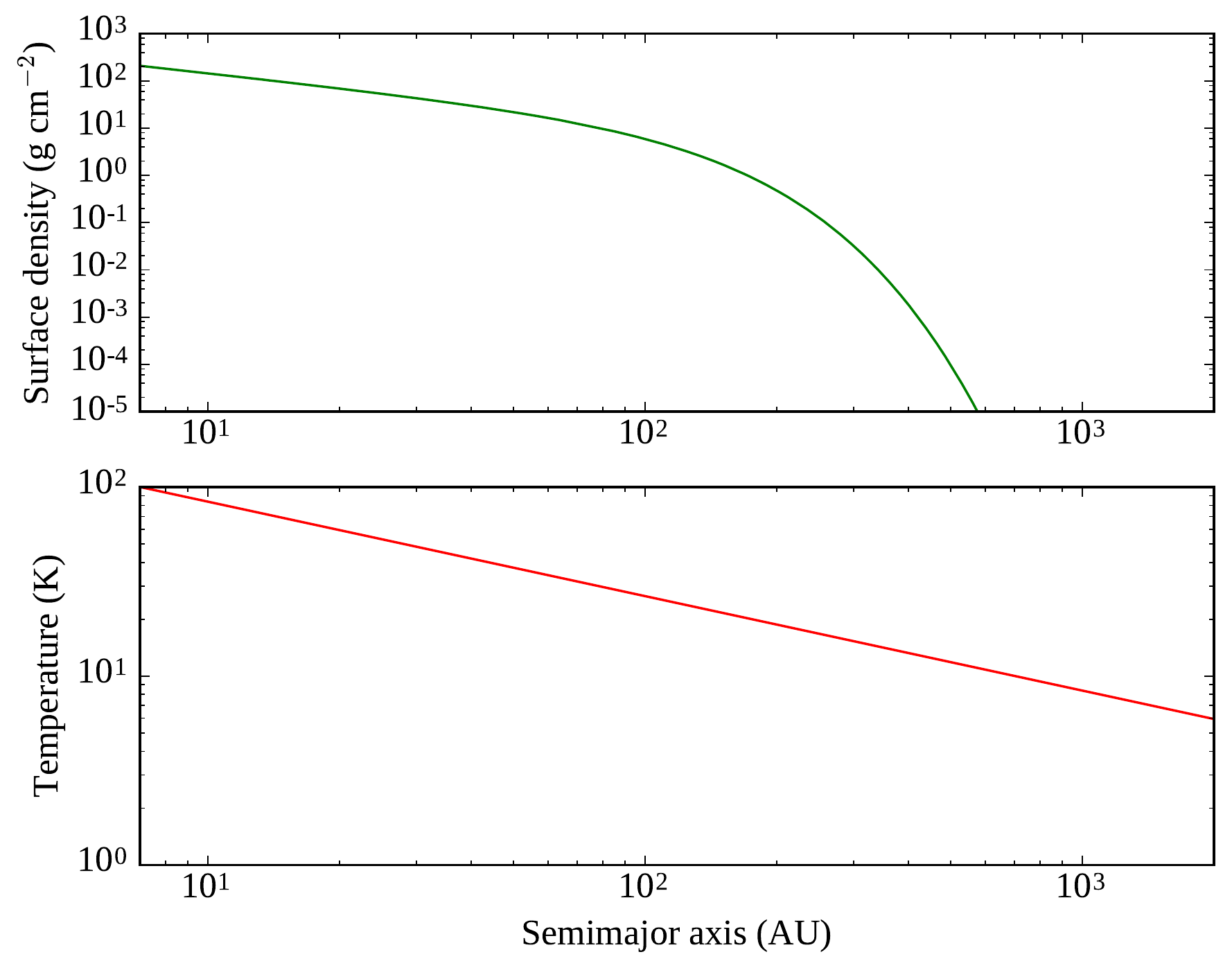}
    \caption{Initial surface density profile (top) and temperature profile (bottom) shared by all our runs. The gas and dust density and temperature structure corresponding to the instantaneous surface density profiles $\Sigma(r,t)$ are calculated via a 1+1D thermochemical model to compute the photoevaporation rate \citep[see][]{Gorti_2015}.}
    \label{fig:initial-profile}
\end{figure}

Figure \ref{fig:initial-profile} shows the initial surface density and temperature profile shared by all our runs. The surface density profile is consistent with observed disks \citep[e.g.][figure 2]{Williams_2011}. Note that the disk is has a large radial extent (beyond 1,000 au) but the surface density drops exponentially after 100 au. Because of the low surface density, the disk is gravitationally stable. The Toomre Q parameter reaches a minimum of $Q = 6$ at around 72 au, so that the stability criterion ($Q > 1$) is met everywhere. The initial dust distribution follows the gas density profile with a dust-to-gas ratio of $Z = 0.01$ and a local size distribution following the prescription of \citet{Birnstiel_2011}.

We performed seven simulations in which we explore how different physical processes impact the formation of planetesimals. Our most complete model is Model \texttt{IAB}. The models are summarized in Table \ref{tab:params}. Here we describe the seven models in more detail:

\begin{itemize}
\item Model \texttt{NP} (``no planetesimals'') is a disk evolution model \textit{without} planetesimal formation. The model is identical to that of \citet{Gorti_2015} and includes viscous accretion, photoevaporation, and dust evolution.

\item Model \texttt{SI} (``streaming instability'') adds planetesimal formation by the streaming instability to Model \texttt{NP} using the results of \citet{Carrera_2015}.

\item Model \texttt{I} (``ice line'') adds the water ice line including condensation, sublimation, and diffusion of water vapor.

\item Model \texttt{IU} (``\texttt{I} + $\uf$'') is a duplicate of Model \texttt{I} with the fragmentation speed reduced to $\uf = 1 \ms$. This is a test to determine whether the simulation results are sensitive to $\uf$.

\item Model \texttt{IB} (``\texttt{I} + bouncing'') includes the water ice line and the bouncing barrier for silicate grains inside the ice line \citep{Guttler_2010}.

\item Model \texttt{IA} (``\texttt{I} + $\at$'') includes the water ice line and a reduced midplane turbulence of $\at = 10^{-4}$ which is more consistent with a dead zone \citep{Oishi_2007}, and is consistent with ALMA observations of outer disks \citep{Flaherty_2015}.

\item Model \texttt{IAB} (``\texttt{I} + $\at$ + bouncing'') is our most complete model. It includes the water ice line, the low midplane turbulence, and bouncing barrier.
\end{itemize}

\begin{table}[h!]
    \caption{Model parameters}
	\label{tab:params}
	\begin{tabular}{lccccr}
		Model & SI & Ice line & Bouncing & $\at$ & $\uf$\\
		\hline
        \texttt{NP}    & No  & No  & No  & $10^{-2}$ & \blue{10 $\ms$} \\
        \texttt{SI}    & \blue{Yes} & No  & No  & $10^{-2}$ & \blue{10 $\ms$} \\
        \texttt{I}     & \blue{Yes} & \blue{Yes} & No  & $10^{-2}$ & \blue{10 $\ms$} \\
        \texttt{IU}    & \blue{Yes} & \blue{Yes} & No  & $10^{-2}$ &  1 $\ms$ \\
        \texttt{IB}    & \blue{Yes} & \blue{Yes} & \blue{Yes} & $10^{-2}$ & \blue{10 $\ms$} \\
        \texttt{IA}    & \blue{Yes} & \blue{Yes} & No  & \blue{$10^{-4}$} & \blue{10 $\ms$} \\
        \texttt{IAB}   & \blue{Yes} & \blue{Yes} & \blue{Yes}  & \blue{$10^{-4}$} & \blue{10 $\ms$} \\
        \hline
	\end{tabular}
    \textbf{Note.} ``SI'' and ``Bouncing'' denote the streaming instability and bouncing barriers. The midplane turbulence parameter is $\at$, and the fragmentation speed is $\uf$. The more physical parameters are marked in blue.
\end{table}

%-----------------------------------------------------------------------------%
% MODEL NP
%-----------------------------------------------------------------------------%
\subsection{Photo-evaporating disk with dust evolution}
\label{sec:methods:NP}

Model \texttt{NP} is a reproduction of the disk evolution model used by \citet{Gorti_2015} and is described here for completeness. At its core, it is the radial one-dimensional (1D) time-dependent viscous evolution model as described by \citet{Lynden-Bell_1974}, with an additional sink term that describes the mass loss due to photo-evaporation,

\begin{equation}\label{eqn:viscous_evolution}
	\frac{\partial \Sigmag}{\partial t} = 
    		\frac{3}{r} \frac{\partial}{\partial r}
            \left[
            	\sqrt{r} \frac{\partial}{\partial r}
                \left( \nu \Sigmag \sqrt{r} \right)
            \right]
            -
            \dot{\Sigma}_\pe(r,t),
\end{equation}
where $r$ is the radial coordinate, $\nu$ is the viscosity, $\Sigmag$ is the gas surface density, and $\dot{\Sigma}_\pe$ is the instantaneous photo-evaporation rate. The viscosity term $\nu$ follows the usual $\alpha$-model of \citet{Shakura_1976},

\begin{equation}\label{eqn:nu}
	\nu = \av \cs^2 / \Omegak,
\end{equation}
where $\cs$ is the local sound speed and $\Omegak$ is the Keplerian frequency. As discussed earlier, the viscous $\av$ may differ from the midplane turbulence $\at$. We use the instantaneous accretion rate onto the star to compute the accretion component of the FUV flux, which we add to the chromospheric flux. We assume that the accretion shock has a black-body spectrum with a temperature of 9,000K \citep{Calvet_1998} and compute the FUV spectrum accordingly. At that temperature, around $4\%$ of the accretion luminosity is in the FUV band (91.2 -- 200 nm), so we estimate

\begin{equation}
	L_{\rm FUV, acc} \approx 0.04
    		\left( \frac{0.8 \, G \, M_\star \, \dot{M}}{R_\star} \right),
\end{equation}
where $M_\star$ and $R_\star$ are the stellar mass and radius, and $\dot{M}$ is the instantaneous accretion rate obtained from the solution to Equation (\ref{eqn:viscous_evolution}). We assume that the chromospheric luminosity in FUV and EUV is of the same magnitude as the X-ray luminosity, which is $L_X \sim 10^{30} \ergs$ for $M_\star = 1 M_\odot$ \citep{Falccomio_2003}.

To compute $\dot{\Sigma}_\pe$ at each time step we build a 1+1D model to resolve the vertical structure of the disk. Photons from the star irradiate the disk surface to heat the dust and gas. Thermal and hydrostatic equilibrium yield the disk structure including the gas temperature $T$ and number density of gas and dust as a function of position $(r,z)$. Following \citet{Gorti_2015}, we use the analytical approximations of \citet{Adams_2004} to compute the evaporation rate $\dot{\Sigma}_\pe$ at each point $(r,z)$. We consider the photoevaporative flow to be launched from the height $z$ where $\dot{\Sigma}_\pe$ is greatest. We refer the reader to \citet{Gorti_2015} for further details.

The dust surface density is evolved independently of the gas component. The dust component is divided into 10 logarithmically spaced size bins ranging from 5 nm to 1 cm. The initial total dust-to-gas ratio is 0.01 across the entire disk. The time evolution of each dust bin $\Sigmap^i$ with particle size $a^i$ is given by the rate of radial mass advection and diffusion. The diffusive flux itself is proportional to the gradient in the mass density ratio of solids to gas. This gives the relation

\begin{equation}\label{eqn:dust-diffusion}
	\frac{\partial \Sigmap^i}{\partial t}
    	= \frac{1}{r} \frac{\partial}{\partial r}
        \left[
        	r \Sigmag D^i \frac{\partial}{\partial r}
            \frac{\Sigmap^i}{\Sigmag}
            -
            r \Sigmap^i u^i
        \right],
\end{equation}
where $u^i$  is the total radial velocity of the dust due to gas drag and the radial pressure gradient, $D^i = \nu / (1 + \St_i^2)$ is the dust diffusivity, and $\St$ is the Stokes number. The evolution of grain sizes follows the prescription in Section 5.2 of \citet{Birnstiel_2011}, with the effects of radial drift added from \citet{Birnstiel_2012}. At every given $r$ and time, grains could drift before they grow (Equation (\ref{eqn:Stdrift})), and drift could result in fragmentation if drift velocities are higher than the threshold (Equation (\ref{eqn:St_df})). Following the notation of \citet{Birnstiel_2012}, we call these limits $a_{\rm drift}$ and $a_{\rm df}$. We then compute

\begin{equation}\label{eqn:a_max}
	a_1 = \min(a_{\rm frag}, a_{\rm df}, a_{\rm drift}).
\end{equation}

Once the maximum particle size $a_1$ is computed, the dust mass is distributed across the size bins following the prescription of \citet{Birnstiel_2011}. All our models follow the same basic prescription. Most models use fragmentation speed $\uf = 10 \ms$ and $\at = 10^{-2}$, but Model \texttt{IU} uses $\uf = 1 \ms$, and Models \texttt{IA} and \texttt{IAB} use $\at = 10^{-4}$. In addition, Models \texttt{IB} and \texttt{IAB} add a maximum size limit due to the bouncing barrier to Equation (\ref{eqn:a_max}).

%-----------------------------------------------------------------------------%
% MODEL SI
%-----------------------------------------------------------------------------%
\subsection{Streaming instability}
\label{sec:methods:SI}

In Model \texttt{SI} and later we instantaneously convert dust into planetesimals whenever the conditions for the streaming instability are met. This is a reasonable approximation because the streaming instability and gravitational collapse of particle clouds occurs on the timescale of just a few orbits \citep[e.g.][]{Johansen_2007}.

We do not track the evolution of planetesimals after they form, but we record the planetesimal mass produced. To trigger the streaming instability we have two requirements. The first requirement is that $\rhop/\rhog > 1$ at the midplane, as this threshold marks the transition to high analytical growth rates by the streaming instability \citep{Youdin_2005}. The value of $\rhop/\rhog$ is given by the particle size and diffusion coefficient $\delta$,

\begin{equation}\label{eqn:midplane-criterion}
\frac{\rhop}{\rhog} = Z \sqrt{\frac{\St + \delta}{\delta}},
\end{equation}
where $Z = \Sigmap / \Sigmag$. In the case of turbulence driven by the magneto-rotational instability, $\delta \approx \at$, but the value may be different if, for example, accretion is driven by disk winds \citep[e.g.][]{Johansen_2014,Johansen_2005,Bai_2013}. The condition $\rhop/\rhog > 1$ can be thought of as a minimum $Z$ or a maximum value for the midplane turbulence $\at$,

\begin{eqnarray}
	\at &<& \frac{Z^2 \St}{1 + Z^2} \label{eqn:delta_max},\\
	Z &>& Z_1 \equiv \sqrt{\frac{\alpha_t}{\St + \at}} \label{eqn:Z1}
\end{eqnarray}

In addition, \citet{Carrera_2015} found a minimum threshold in Z for dense particle filaments to form by the streaming instability; this threshold occurs even when $\at$ is negligible. We computed a least-squares fit for the results of \citet{Carrera_2015} and obtained the condition,

\begin{equation}\label{eqn:Z2}
    Z > Z_2 \equiv 10^{-1.86 + 0.3 \left(0.98 + \log_{\rm 10}\St\right)^2}.
\end{equation}

\begin{figure}[t!]
	\includegraphics[width=\columnwidth]{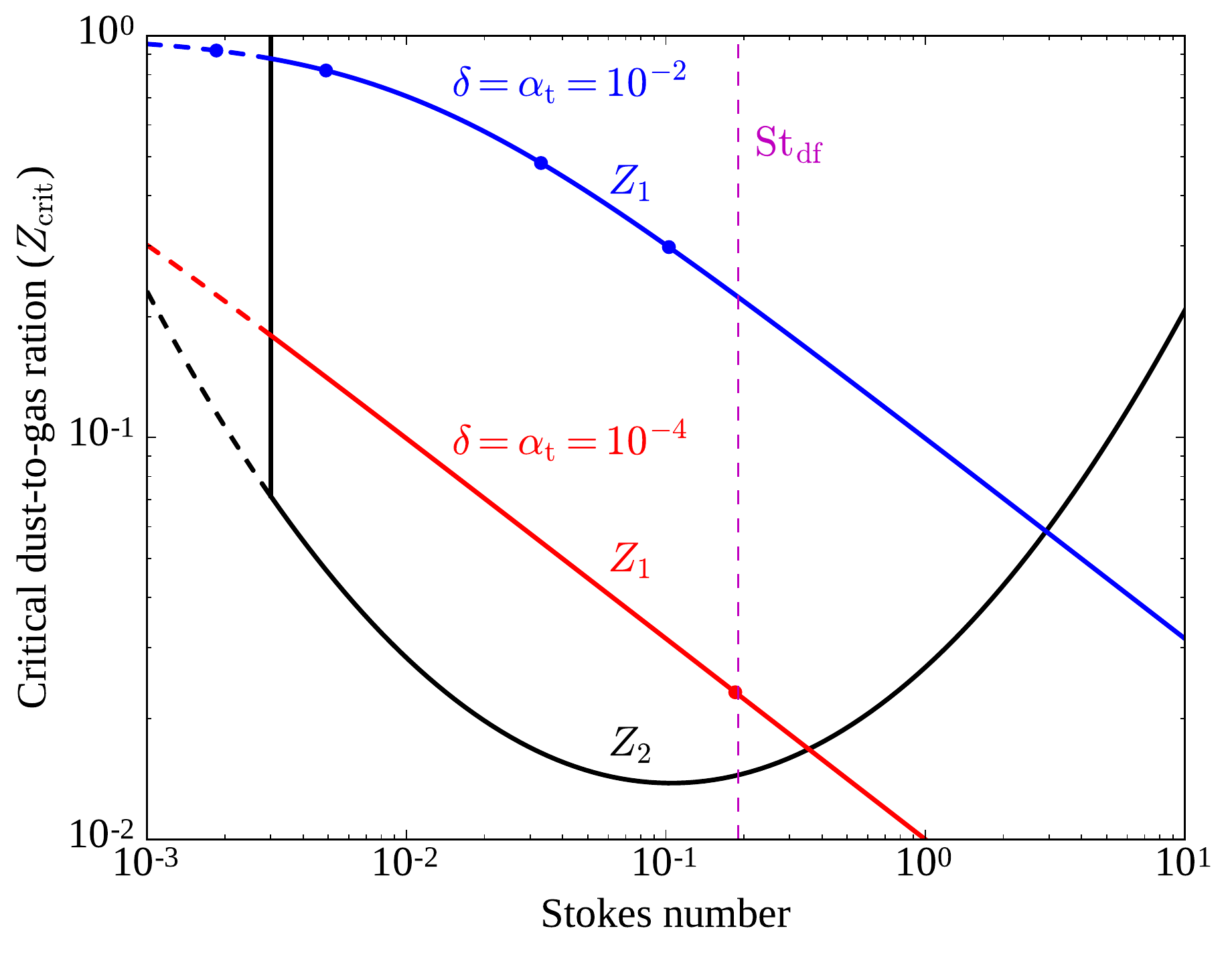} \\
	\includegraphics[width=\columnwidth]{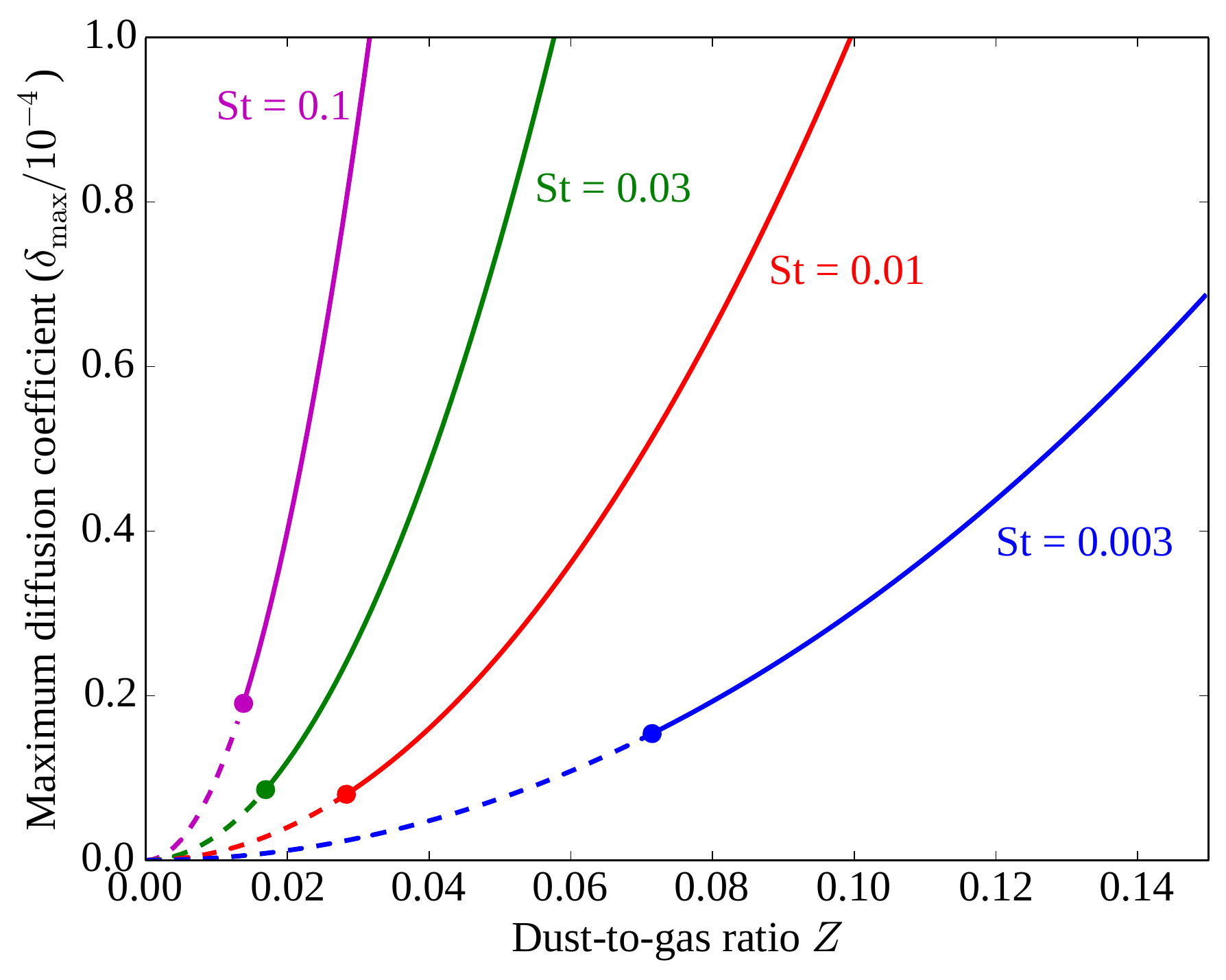}
    \caption{\textit{Top}: Minimum dust-to-gas ratio ($Z$) needed for planetesimal formation. The black curve is the least-squares fit for the results of \citet{Carrera_2015} ($Z_2$, Equation (\ref{eqn:Z2})). The red and blue curves mark the $Z$ needed for $\rhop > \rhog$ on the midplane ($Z_1$, Equation (\ref{eqn:Z1})) assuming that the diffusion coefficient equals the midplane turbulence ($\delta = \at$) for two values of $\at$. For $\at = 10^{-2}$ (blue), we mark the $\St_{\rm frag}$ at 0.1, 1, 10, and 100 au (see points left-to-right) using the temperature profile of Model \texttt{SI}; for $\at = 10^{-4}$ (red) only $\St_{\rm frag}$ at 0.1 au is visible in the plot. The vertical dashed line is the drift-fragmentation limit (Equation (\ref{eqn:St_df})) which is nearly constant across the disk. \textit{Bottom:} Maximum diffusion coefficient (Equation (\ref{eqn:delta_max})) as a function of $Z$ for four sample particle sizes. The solid lines correspond to $Z > \Zcrit$, and dashed lines are $Z < \Zcrit$. For the conditions expected in the disk, the requirement that $\rhop/\rhog > 1$ is more stringent than $Z > \Zcrit$.}
    \label{fig:Zcrit}
\end{figure}

In line with the results of \citet{Carrera_2015}, we also require $\St \ge 0.003$. More recent work by \citet{Yang_2016} shows that the streaming instability can be effective for even smaller particles. We will incorporate those new results in a future work. Figure \ref{fig:Zcrit} shows both criteria together. Let $Z_1$ and $Z_2$ be right-hand-side of Equations (\ref{eqn:Z1}) and (\ref{eqn:Z2}) respectively. Together, the conditions for the streaming instability become

\begin{eqnarray}
	Z &>& \Zcrit = \max(Z_1, Z_2) \label{eqn:Zcrit} \\
    \St &>& 0.003
\end{eqnarray}

The simulations conducted by \citet{Carrera_2015} always had a uniform particle size. In our simulations we have to generalize this result to a situation where there is a range of particle size bins. To do this we start by computing the value of $Z$ for the largest size bin. If $Z \ge \Zcrit$ , we remove 90\% of the mass in that bin. If $Z < \Zcrit$, we compute $Z$ for the top two bins, and re-compute $\Zcrit$ as if the entire dust mass was in the lower mass bin. We repeat this process until planetesimals form, or we reach the minimum particle size of $\St~=~0.003$. This is a conservative choice because it discounts the effect of larger grains, even though they tend to dominate the streaming instability \citep{Bai_2010}.

We assume that planetesimals always form when the streaming instability is active. We believe that this assumption is justified because the streaming instability gathers solids into long-lived stable filaments. If the filaments do not form planetesimals immediately, they will become effective traps for other solid particles drifting radially from the outer disk. Therefore, the filament density will grow on the radial drift timescale. The filament density will drastically increase when photoevaporation dissipates the last remaining gas from the disk. In section \ref{sec:other:roche} we discuss the physics of gravitational collapse in greater detail.

%-----------------------------------------------------------------------------%
% MODEL I
%-----------------------------------------------------------------------------%
\subsection{Water ice line}
\label{sec:methods:I}

Model \texttt{I} and later include the water ice line. Near the ice line we must model the sublimation and condensation of water ice onto solid grains. We use a pure thermal desorption/absorption process, meaning that ice forms whenever the partial pressure of water vapor is greater than the saturation pressure. We ignore photodesorption (e.g.~from UV) because we are only interested in the formation of ices on the midplane, which is shielded from high-energy photons. The saturation pressure of water is given by

\begin{equation}\label{eqn:Psat}
	P_{\rm sat} =	2.53 \times 10^{13}
    				\exp \left( \frac{- 6070 \K}{T} \right)
                	\dyne \; \cm^{-2}
\end{equation}
where $T$ is the local gas temperature \citep{Leger_1985}. The partial pressure of water is

\begin{equation}\label{eqn:Pwater}
	P_{\rm H_2O} =  n_{\rm H_2O} \kb T =  n_{\rm H_2} X_{\rm H_2O} \kb T,
\end{equation}
where $n_{\rm H_2}$ and $n_{\rm H_2O}$ are the number densities of hydrogen and water, $\kb$ is the Boltzmann constant and $X_{\rm H_2O}$ is the water abundance. Ice forms whenever $P_{\rm H_2O} > P_{\rm sat}$, which simplifies to

\begin{equation}\label{eqn:nH2}
	n_{\rm H_2} > \frac{2.53 \times 10^{13} \; \dyne \; \cm^{-2}}{X_{\rm H_2O} \; \kb T}
    			\exp \left( \frac{- 6070 \K}{T} \right).
\end{equation}

The run begins with $X_{\rm H_2O} = 3 \times 10^{-4}$ everywhere, which corresponds to a water mass fraction of 0.005. Note that the CO ice line can also be important for disks, but is ignored here. Our chosen water abundance therefore reflects all the oxygen being bound in water ice and slightly underestimates the total solid mass in the outer disk beyond the putative CO ice line. For simplicity, we further assume that the water is either all vapor or all ice, and we use the midplane temperature with mean number density $\langle n_{\rm H_2} \rangle
=\langle \rho_{\rm H_2} \rangle/m_{\rm H_2}$ integrated over one scale height.

When $\langle n_{\rm H_2} \rangle$ satisfies Equation (\ref{eqn:nH2}), all the water is added to the dust component uniformly, with constant d$m/m$ across all grain size bins. This differs from the approach of \citet{Ros_2013}, who kept d$a$ constant. This latter choice is consistent with pure vapor condensation, but in a disk where dust particles have frequent collisions that result in coagulation and fragmentation, the water component could be quickly distributed across bin sizes.

We keep track separately of $\Sigma_{\rm ice}$ and $\Sigma_{\rm vapor}$ in order to determine the water fraction in dust or gas. Inside the ice line, $\Sigma_{\rm ice}$ is locally released into $\Sigma_{\rm vapor}$ which advects and diffuses like the rest of the disk gas.

%-----------------------------------------------------------------------------%
% MODEL IB
%-----------------------------------------------------------------------------%
\subsection{Bouncing barrier}
\label{sec:methods:IB}

Inside the water ice line, the solid component is made entirely of silicate grains which may experience a bouncing barrier \citep{Zsom_2010,Guttler_2010}. Models \texttt{IB} and \texttt{IAB} implement this bouncing barrier. To do this we manually fit a straight line through the plot in Figure 11 of \citet{Guttler_2010}. We use the results for collisions between equal-mass compact grains. This is the most conservative choice we can make, as the other choices would produce larger grains that more easily participate in the streaming instability. The choice of compact grains is also consistent with our opacity calculation. Our manual fit to the mass of particles at the bouncing barrier $m_{\rm b}$, is

\begin{equation}\label{eqn:m_bounce}
	\log_{10} \left( \frac{m_{\rm b}}{\rm g} \right)
    = -2.08 \log_{10} \left( \frac{\Delta u_{\rm coll}}{\rm cm \; s^{-1}} \right) - 7.33
\end{equation}

Inside the ice line, we use Equation (\ref{eqn:m_bounce}) to compute the corresponding grain size at the bouncing barrier $a_{\rm b}$ and replace Equation (\ref{eqn:a_max}) with the minimum of $a_1$ and $a_{\rm b}$.

%-----------------------------------------------------------------------------%
% MODEL IA
%-----------------------------------------------------------------------------%
\subsection{Reduced midplane turbulence}
\label{sec:methods:IA}

Most of our models follow a simple $\alpha$-disk prescription \citep{Shakura_1976} in which $\at = \av = 10^{-2}$ everywhere in the disk. But on the midplane and in the outer disk, the actual level of turbulence is likely to be significantly lower. This is important for planetesimal formation because solids are mostly on the midplane and the collision speed between solid particles is proportional to $\sqrt{\at}$ (Equation (\ref{eqn:du_coll})), and because the midplane turbulence sets the amount of sedimentation (set $\delta = \at$ in Equation (\ref{eqn:midplane-criterion})).

First, from around 0.2-1 au to 30-100 au, protoplanetary disks are thought to experience layered accretion, in which the midplane is inside an MRI-inactive ``dead zone'' that experiences very little turbulence \citep{Gammie_1996,Fromang_2002}. Simulations of the magneto-rotational instability by \citet{Oishi_2007} give an $\at$ of $10^{-5}$ inside the dead zone. Alternatively, simulations of the vertical shear instability give $\at = 10^{-4}$ on the midplane \citep{Stoll_2014}. While there are no strong theoretical constraints beyond 100 au, ALMA CO observations indicate a three-sigma upper limit of $\alpha < 10^{-3}$ for the outer regions ($R > 30$ au) of the disk around around HD 163296 \citep{Flaherty_2015}. Recent ALMA observations of the TW Hya disk \citep{Teague_2016} suggest a value of $\alpha$ closer to our adopted default value of 0.01.

Our choice thus far of $\at = 10^{-2}$ in the midplane may therefore be high, and it is possible that true particle collision speeds are in fact lower than calculated. On the other hand, disk lifetimes indicate an effective viscous $\av \sim 10^{-2}$, and lower values yield disk lifetimes far longer than observed \citep{Gorti_2015}. In Models \texttt{IA} and \texttt{IAB} we separate $\av$ and $\at$: $\av$ is used to compute the viscous evolution of the disk, and $\at$ is used to compute the collision speed between solid particles. We kept the viscous evolution at $\av = 10^{-2}$ as in our previous runs, and we chose a conservative $\at = 10^{-4}$. This lowers collision speeds by a factor 10 while maintaining the nominal accretion rate of the gas and giving reasonable disk lifetimes.

%%%%%%%%%%%%%%%%%%%%%%%%%%%%%%%%%%%%%%%%%%%%%%%%%%%%%%%%%%%%%%%%%%%%%%%%%%%%%%%
%
% RESULTS
%
%%%%%%%%%%%%%%%%%%%%%%%%%%%%%%%%%%%%%%%%%%%%%%%%%%%%%%%%%%%%%%%%%%%%%%%%%%%%%%%
\section{Results and discussion}
\label{sec:results}

%-----------------------------------------------------------------------------%
% Planetesimal formation
%-----------------------------------------------------------------------------%
\subsection{Planetesimal formation}
\label{sec:results:SI}

\begin{figure}[ht!]
	\includegraphics[width=0.99\columnwidth]{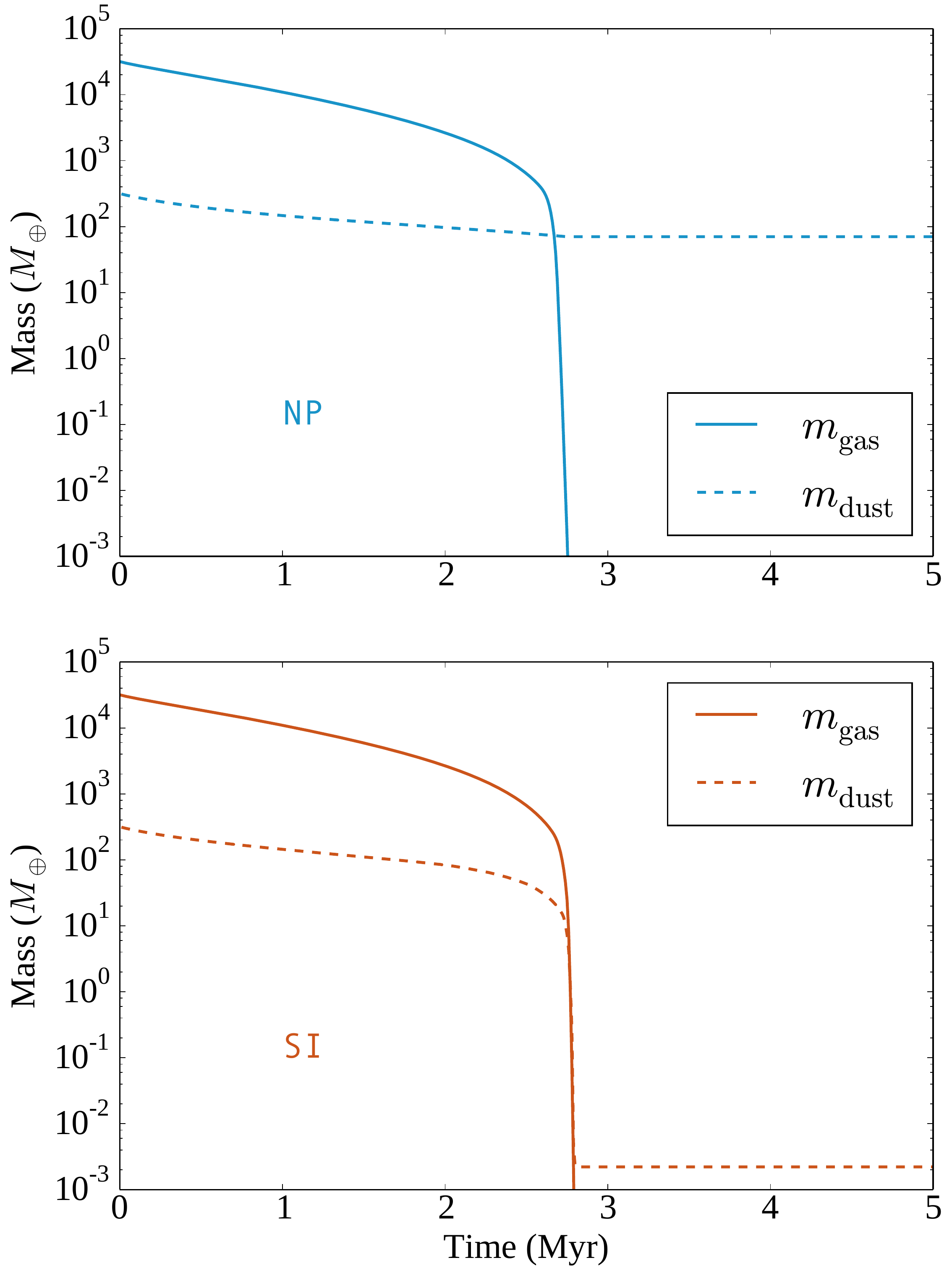}
    \caption{Total mass in gas (solid) and dust (dashed) as a function of time. Model \texttt{NP} (top) is the disk evolution model of \citet{Gorti_2015}. The disk clears after 2.75 Myr and leaves behind 70 $M_\oplus$ of dust. Model \texttt{SI} (bottom) introduces planetesimal formation by the streaming instability. As the dust is converted into planetesimals, this model leaves behind only 0.002 $M_\oplus$ of dust.}
    \label{fig:disk-evo}
\end{figure}

Figure \ref{fig:disk-evo} shows how the formation of planetesimals affects the evolution of a protoplanetary disk. Models \texttt{NP} and \texttt{SI} have the same disk parameters, but Model \texttt{SI} introduces planetesimal formation by the streaming instability as described in section \ref{sec:methods:SI}. The main consequence of planetesimal formation is the depletion of dust. Model \texttt{NP} leaves behind $70 M_\oplus$ of dust. This much dust is inconsistent with observations of debris belts \citep{Wyatt_2008}. In Model \texttt{SI}, only $0.002 M_\oplus$ of dust remain, as the rest is converted into planetesimals. In reality, once the protoplanetary disk evolves into a debris disk, the remaining dust is quickly removed by either photon pressure or Poynting-Robertson drag, and new dust is re-generated through collisional grinding of surviving planetesimals \citep{Wyatt_2008}. We do not attempt to model these processes because the debris phase is beyond the scope of our investigation and not relevant to our goal of understanding planetesimal formation. Therefore, the solid component in Figure \ref{fig:disk-evo} stops evolving once the gas dissipates.

\begin{figure*}[ht!]
	\includegraphics[width=0.99\columnwidth]{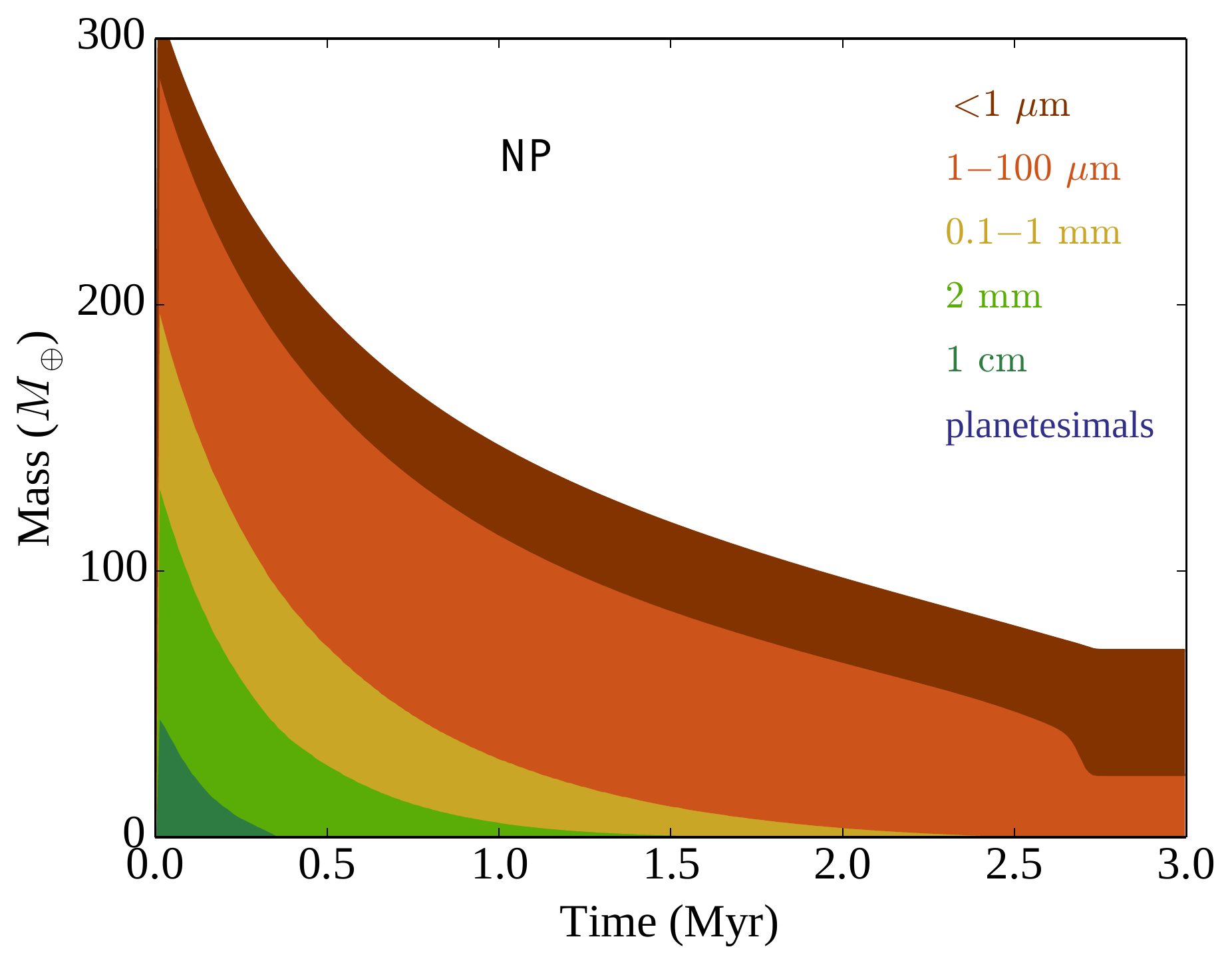}
	\includegraphics[width=0.99\columnwidth]{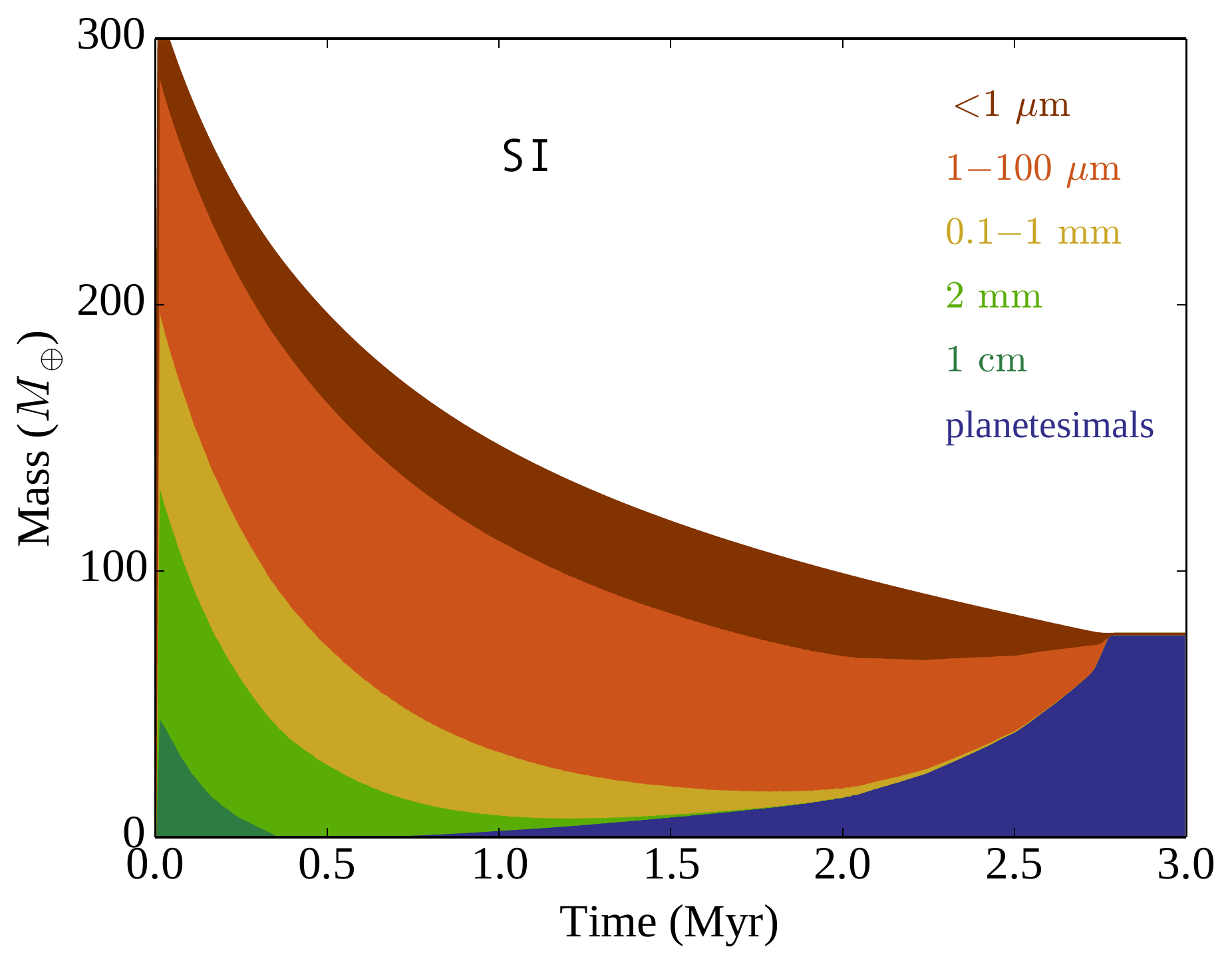}
    \caption{Total mass in planetesimals and in various dust size bins as a function of time. In Model \texttt{NP} (left) there is no mechanism to remove dust grains once photoevaporation forms an inner cavity. This model leaves behind 47 $M_\oplus$ of sub-$\mu$m dust, 23 $M_\oplus$ of 1--100 $\mu$m dust, and trace amounts of larger grains. In Model \texttt{SI} (right), dust is efficiently converted into planetesimals. This model converts 76 $M_\oplus$ of dust into planetesimals, leaving behind only 0.002 $M_\oplus$ of sub-$\mu$m and trace amounts of larger grains.}
    \label{fig:dust-evo}
\end{figure*}

\begin{table*}[ht!]
	\centering
    \caption{Simulation results}
	\label{tab:results}
	\begin{tabular}{rlrrrrrrr}
            & & \multicolumn{7}{c}{Planetesimal mass in radial bins ($M_\oplus$)} \\
      Model & Summary & Total & $<$1 au & 1-3 au & 3-10 au & 10-30 au & 30-100 au & $>$100 au\\
      \hline
      \texttt{SI}  & Streaming instability only &   76.34 &  0.01 &  0.27 &  4.03 &  4.16 &  7.98 &  59.89 \\
      \texttt{I}   & \texttt{SI} + water ice line       &  122.62 &  0.02 &  0.32 &  6.78 &  9.03 & 12.78 &  93.70 \\
      \texttt{IU}  & \texttt{I} + reduced $u_{\rm frag}$&  153.08 &  0.01 &  0.17 &  5.09 &  7.01 & 11.92 & 128.87 \\
      \texttt{IB}  & \texttt{I} + bouncing barrier      &  122.98 &  0.00 &  0.08 &  5.26 & 10.84 & 13.12 &  93.67 \\
      \texttt{IA}  & \texttt{I} + reduced $\at$         &   74.52 &  0.08 &  0.15 &  0.16 &  0.05 &  0.16 &  73.92 \\
      \texttt{IAB} & \texttt{IAB} + bouncing barrier     &   83.58 &  1.03 &  7.35 &  1.62 &  0.44 &  0.18 &  72.95 \\
      \hline
	\end{tabular} \\
    \textbf{Note.} $\at$ is midplane turbulence. It sets the collision speed (Equation (\ref{eqn:du_coll})) and midplane density ($\delta = \at$ in Equation (\ref{eqn:midplane-criterion})).
\end{table*}

\begin{figure*}[ht!]
	\centering
	\includegraphics[width=1.98\columnwidth]{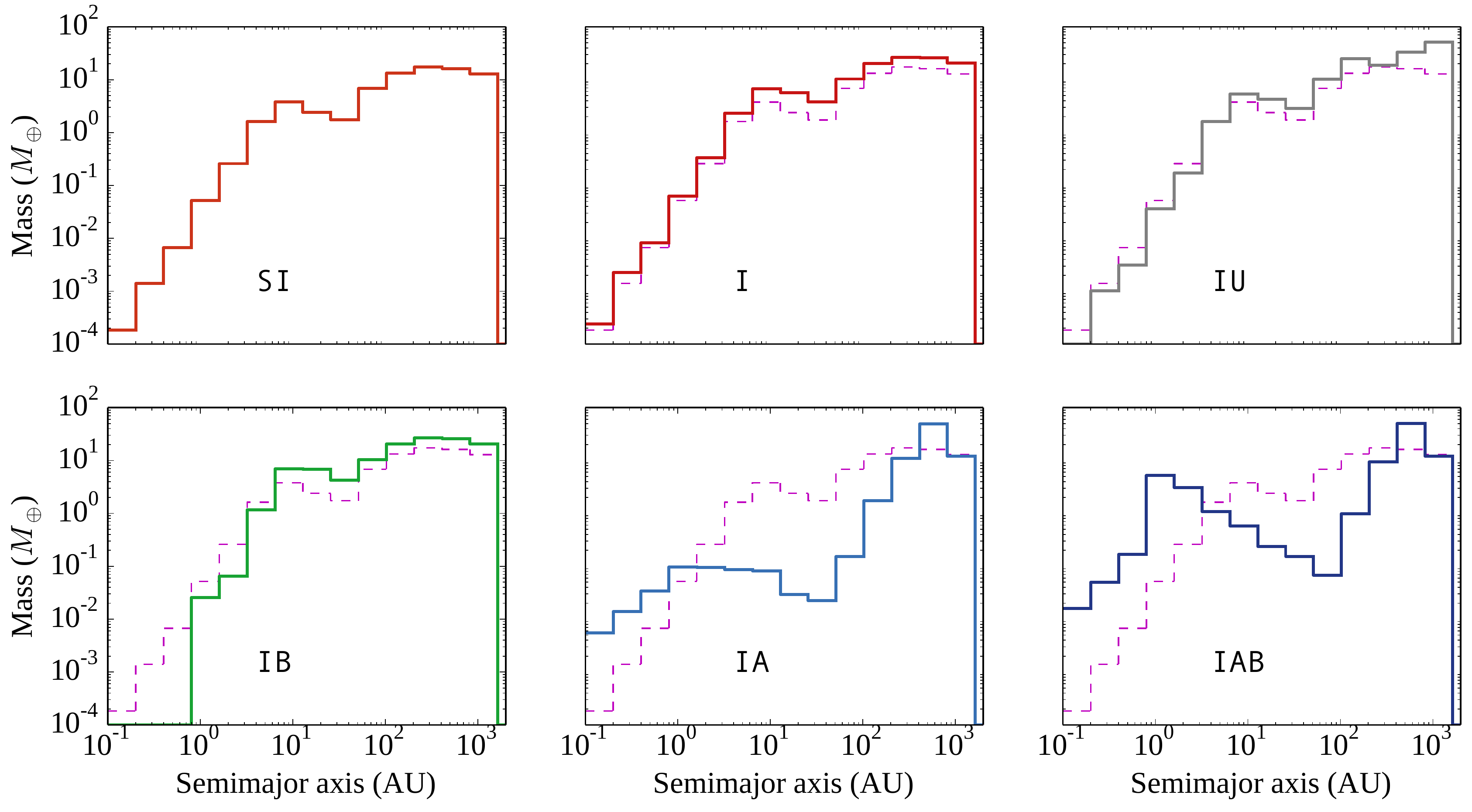}
    \caption{Total planetesimal mass produced in all the models. Each radial bin corresponds to a factor of two in semimajor axis. As a point of reference, the histogram for Model \texttt{SI} is shown as a dashed line on top of the other models. There is little difference between Models \texttt{SI}, \texttt{I}, and \texttt{IU}. However, introducing the bouncing barrier (\texttt{IB}, \texttt{IAB}) and low midplane turbulence (\texttt{IA}, \texttt{IAB}) has a large effect on the distribution of planetesimals.}
    \label{fig:histogram-all}
\end{figure*}

Figure \ref{fig:dust-evo} shows the fate of the dust component in greater detail. Model \texttt{NP} finishes with $23 M_\oplus$ of 1-100$\mu$m grains, and $47 M_\oplus$ of sub-$\mu$m grains. However, with the inclusion of the streaming instability in Model \texttt{SI}, $76 M_\oplus$ of dust are converted into planetesimals and only $0.002 M_\oplus$ of sub-$\mu$m grains survive to the end, along with negligible amounts of larger grains.

\begin{figure}[h!]
	\includegraphics[width=\columnwidth]{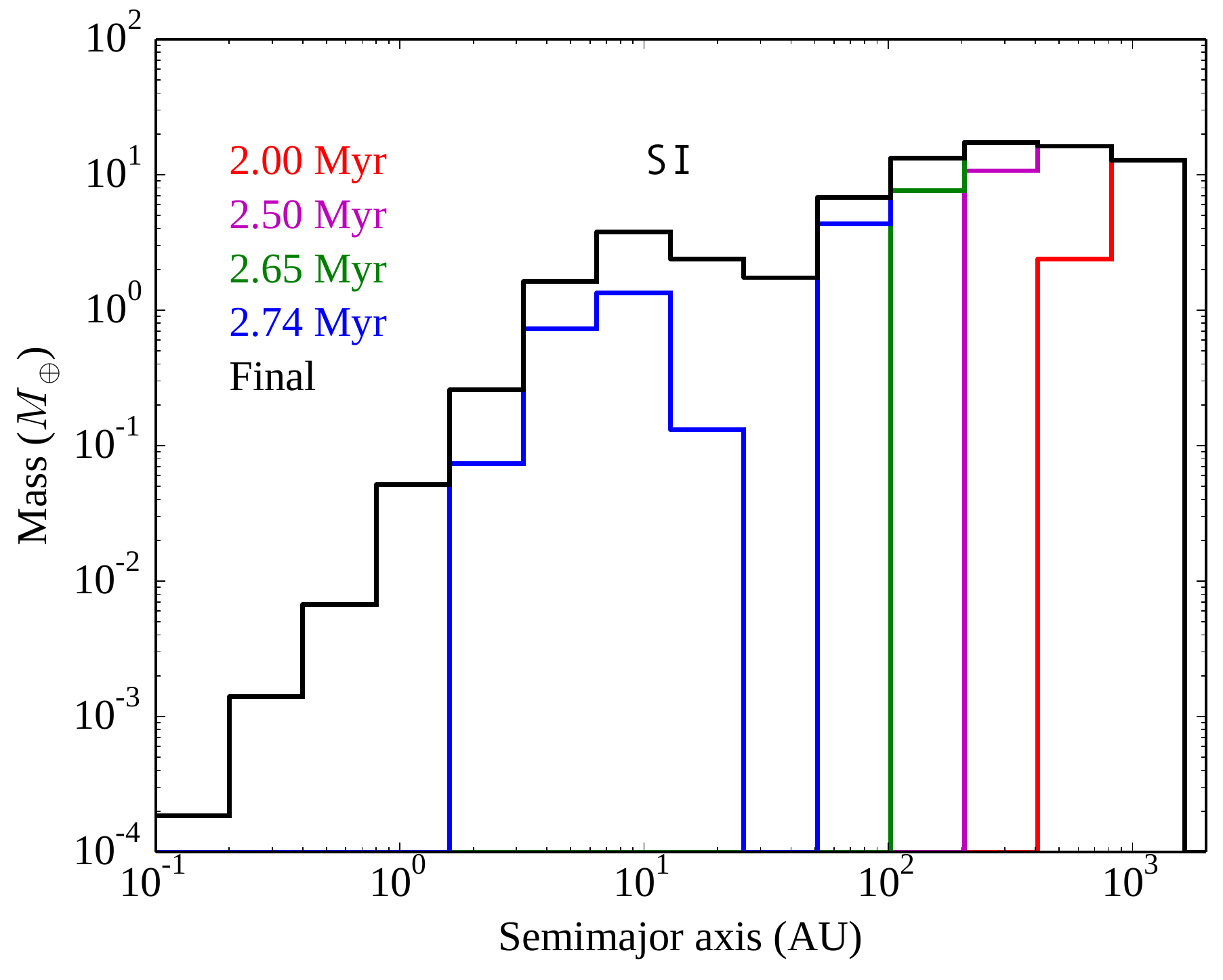}
    \caption{Total planetesimal mass produced in Model \texttt{SI} at different points in time. Each radial bin corresponds to a factor of two in semimajor axis. Although this run produces $76 M_\oplus$ of planetesimals, only $0.29 M_\oplus$ form in the inner 3 au (see Table \ref{tab:results}). Planetesimals initially form only in the outer disk. As the disk dissipates, the planetesimal formation region moves inward. When photoevaporation carves a gap in the inner disk, a second wave of planetesimal formation appears.}
    \label{fig:histogram-SI}
\end{figure}

In Model \texttt{SI}, about 24\% of the original dust in the disk is converted into planetesimals. However, very few of those planetesimals form in the inner disk. Table \ref{tab:results} and Figure \ref{fig:histogram-all} show the final distribution of planetesimals for all the runs. In all cases, most of the planetesimals form beyond 100 au, which indicates the presence of a massive outer debris belt. We note that our debris belt is at larger radii than typically observed \citep{Wyatt_2008}. Given that we consider neither migration of planetesimals after formation nor the ensuing stages of planet formation, we cannot meaningfully compare our results to planetary system architectures.

\begin{figure*}
	\centering
	\includegraphics[width=1.9\columnwidth]{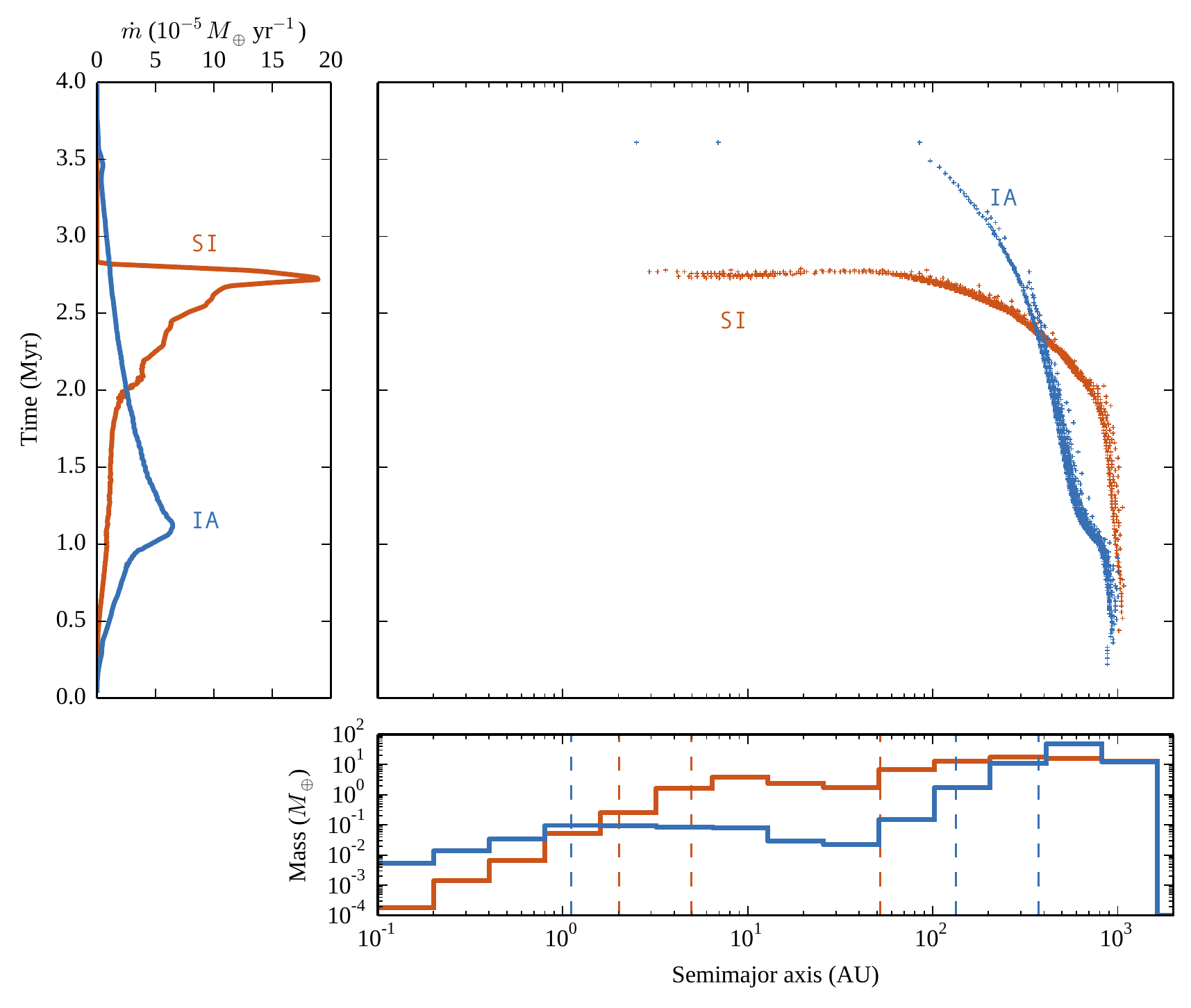}
    \caption{Time and place where planetesimals form in models \texttt{SI} (orange) and \texttt{IA} (blue). Models \texttt{I}, \texttt{IU}, and \texttt{IB} look similar to \texttt{SI} and Model \texttt{IAB} is shown in Figure \ref{fig:killer-SI-IAB}. In Model \texttt{IA} the midplane turbulence is reduced to $\at = 10^{-4}$. This leads to the formation of larger grains which reduces FUV photoevaporation, which prolongs the disk lifetime by approximately 1 Myr. The main plot is a spacetime diagram where each point represents $0.1 M_\oplus$ of planetesimals. The bottom plot shows the total planetesimal mass in radial bins, where each bin represents a factor of two in radius (see also Figure \ref{fig:histogram-SI} and Table \ref{tab:results}). The vertical lines mark the points on the disk that produce 0.1, 1, and 10 $M_\oplus$. The left plot shows the planetesimal formation rate over time.}
    \label{fig:killer-SI-IA}
\end{figure*}

Figures \ref{fig:histogram-SI} and \ref{fig:killer-SI-IA} give a general picture of how planetesimal formation occurs in Model \texttt{SI}. Planetesimals always begin to form early (after $\sim$500,000 years) in the outer disk. Even in the absence of photoevaporation, gas viscously spreads out to larger radii while dust generally drifts in radially. This causes an enhancement in the local dust-to-gas ratio in the outer disk. With photoevaporation, gas is preferentially removed from the outer disk, increasing $Z$ above $Z_{\rm crit}$ and triggering the streaming instability. FUV-driven photoevaporation begins in the outer disk at early times, and this is where planetesimals first form in all our models. As the disk evolves, its outer radius decreases due to FUV photoevaporation and planetesimal formation moves inward. 

In Model \texttt{SI}, there is not much mass loss due to photoevaporation in the inner disk at early epochs. Viscous accretion depletes both gas and dust, and there is no preferential removal of gas at small radii to bring $Z$ above $Z_{\rm crit}$. At later epochs, when the surface density of gas decreases, the mass accretion rate eventually falls below the photoevaporation rate and a gap opens in the disk \citep[as shown previously by][]{Alexander_2009}. The disk dissipates rapidly after this, increasing the solids-to-gas mass ratio globally. Model \texttt{SI} shows a last-minute strong burst of planetesimal formation which leaves behind very little remaining dust. Models \texttt{I}, \texttt{IU}, and \texttt{IB} look very similar to \texttt{SI}, and show all the same features. Models \texttt{IA} and \texttt{IAB} show different evolution and will be discussed in later sections.

It is important to note that the terrestrial planets of the solar system are water poor, and could not have formed from planetesimals that originated beyond 2-3 au (approximate location of the ice line). Therefore, the total planetesimal mass that forms within 3 au is an important benchmark for any planetesimal formation model. We will refer to the inner 3 au as the terrestrial zone. Model \texttt{SI} does not produce enough planetesimals in this region to form the terrestrial planets in the solar system (Table \ref{tab:results}).

%-----------------------------------------------------------------------------%
% Water ice line
%-----------------------------------------------------------------------------%
\subsection{Water ice line}
\label{sec:results:I}

Model \texttt{I} (Table \ref{tab:params}) has the same model parameters as Model \texttt{SI}, but the addition of a water ice line which probably plays an important role in planetesimal formation. Beyond the ice line, water contributes to the solid component of the disk -- Model \texttt{I} starts with 476 $M_\oplus$ of solids, compared to only 313 $M_\oplus$ for Model \texttt{SI}. Water ice maintains a high fragmentation barrier of $\uf \sim 10 \ms$ \citep{Wada_2009}, so that ice-silicate aggregates can achieve larger Stokes numbers than pure silicates. Finally, water ice may also play an important role in transporting solids to the inner disk because the larger Stokes numbers of ice-silicate aggregates are more susceptible to radial drift (Equation (\ref{eqn:u_drift})). But when the aggregate crosses the ice line, the ice component will sublimate and leave behind the silicate dust. Since pure silicate aggregates cannot achieve the same Stokes numbers as the ice-rich aggregates, they experience much lower radial drift, so they would tend to accumulate. The greater solid mass, and the increase in the dust-to-gas ratio could contribute to the formation of planetesimals in the terrestrial zone. We explore this possibility in Model \texttt{I}.

Table \ref{tab:results} shows that Model \texttt{I} does produce significantly more planetesimals than Model \texttt{SI} (123 $M_\oplus$ vs 76 $M_\oplus$) as would be expected given the larger solid mass. However, most of the gains remain in the outer disk. Figure \ref{fig:histogram-all} shows that the distribution of planetesimals in Model \texttt{I} is basically the same as Model \texttt{SI}. Water ice alone is not enough to produce a large planetesimal mass in the terrestrial zone.

%-----------------------------------------------------------------------------%
% Fragmentation barrier
%-----------------------------------------------------------------------------%
\subsection{Fragmentation barrier}
\label{sec:results:IU}

As noted earlier, the $\uf = 10 \ms$ limit is consistent with the ice-dust aggregates that would be found beyond the ice line \citep{Wada_2009,Gundlach_2015}. However, given the uncertainty in disk dust properties, we next test how our results would change if ices fragment more easily than expected. Model \texttt{IU} is a copy of Model \texttt{I}, with the fragmentation speed reduced to $\uf = 1 \ms$. This change increases the total planetesimal mass (153 $M_\oplus$ vs 123 $M_\oplus$). The reason for this is that easier fragmentation leads to smaller grains that do not drift readily. Table \ref{tab:results} shows that the gain in planetesimals occurs entirely beyond 100 au. Closer to the star, the lack of radial drift actually depletes the planetesimals. In the terrestrial zone (inside 3 au), the planetesimal mass drops by 50\%, as more of the solids turn into planetesimals further out.

Figure \ref{fig:histogram-all} again shows that the distribution of planetesimals is essentially the same as that of Models \texttt{SI} and \texttt{I}. There are more planetesimals in the outer disk, especially at 1000 au, and fewer planetesimals in the inner disk, inside 3 au. These results show that the fragmentation speed does not drastically change our results, though a low fragmentation barrier may exacerbate the difficulties in forming dry planetesimals in the terrestrial zone.

%-----------------------------------------------------------------------------%
% Bouncing barrier
%-----------------------------------------------------------------------------%
\subsection{Bouncing barrier}
\label{sec:results:IB}

\begin{figure}[h!]
	\includegraphics[width=\columnwidth]{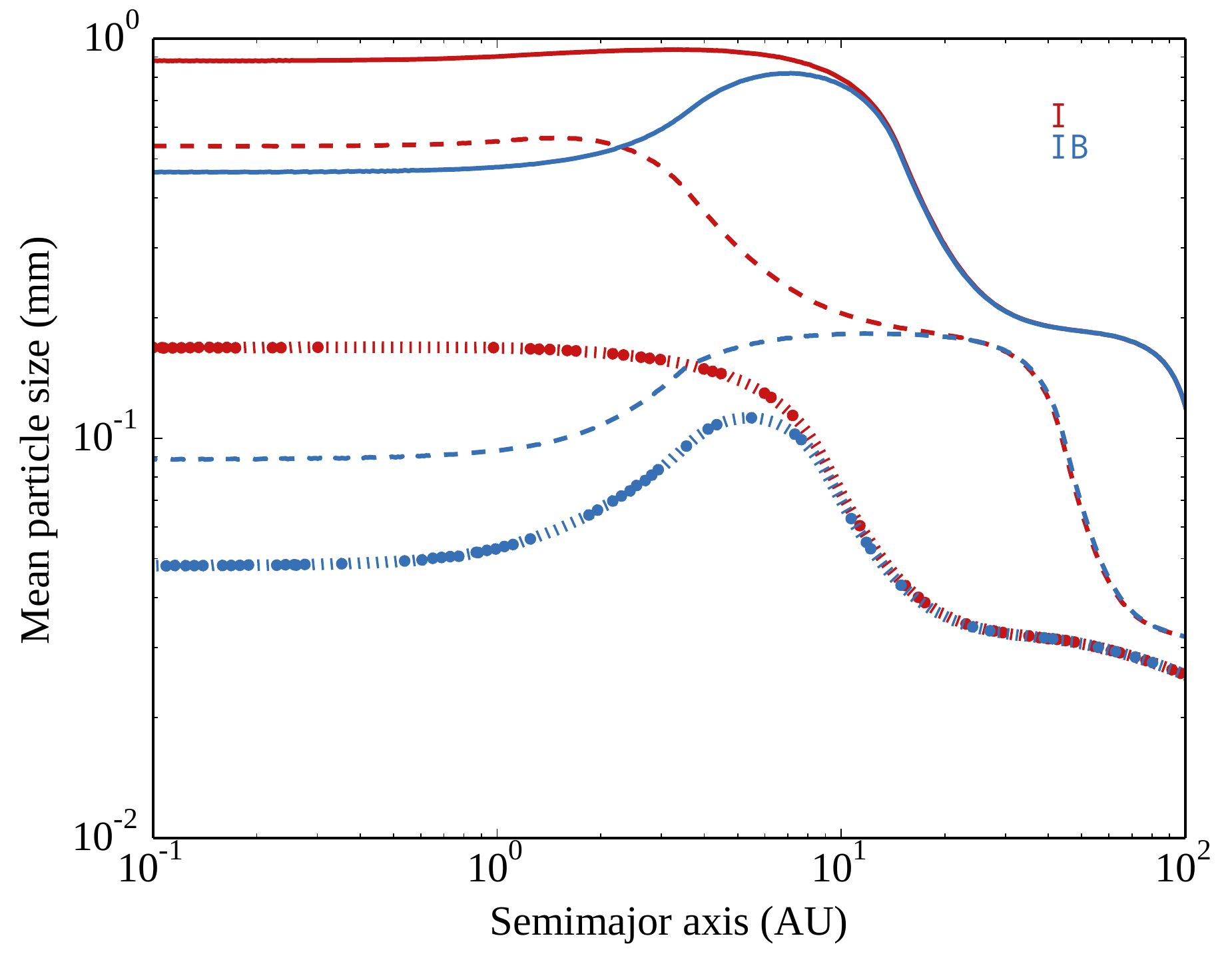}
    \caption{The mass-weighted mean size of dust grains in Models \texttt{I} (red) and \texttt{IB} (blue) as a function of semimajor axis. The snapshots are taken at 1.2 Myr (solid), 1.8 Myr (dashed), and 2.4 Myr (dotted). Inside the ice line ($\sim$3 au) the bouncing barrier keeps the grains small. As the gas dissipates, the particle sizes decrease. The drop in gas density increases the Stokes numbers (Equation (\ref{eqn:St})) which in turn increases the particle collision speeds (Equation (\ref{eqn:du_coll})), leading to smaller particle sizes.}
    \label{fig:mean-a-I-IB}
\end{figure}

Model \texttt{IB} is a copy of Model \texttt{I} (both have a water ice line and $\uf=10 \ms$) with the addition of the bouncing barrier for silicates inside the ice line (Table \ref{tab:params}). Our implementation of the bouncing barrier (section \ref{sec:methods:IB}) is derived from \citet{Guttler_2010}. The bouncing barrier affects pure silicate grains and keeps them from reaching the fragmentation limit. Figure \ref{fig:mean-a-I-IB} shows the mass-weighted mean particle size for Models \texttt{I} and \texttt{IB} at three different points in time. There are three key features in this plot,

\begin{itemize}
\item In the outer disk, Models \texttt{I} and \texttt{IB} look identical. Inside the ice line (approximately 3 au) the bouncing-limited grains of Model \texttt{IB} are significantly smaller. The transition in particle size is smooth and goes between 3 and $\sim$10 au because the small bouncing-limited grains can diffuse outward.

\item The jump in particle size that occurs in the outer disk (at around 20 au) marks the transition from drift-limited growth (outer disk) to fragmentation-limited growth (middle and inner disk).

\item As time passes and the disk dissipates, particles become more strongly coupled to the gas and their collision speeds increase (Equation (\ref{eqn:du_coll})) so the particles must become smaller.
\end{itemize}

Table \ref{tab:results} and Figure \ref{fig:histogram-all} show the effect of the bouncing barrier. Not surprisingly, Model \texttt{IB} looks extremely similar to Model \texttt{I} beyond 3 au. Inside 3 au, the planetesimal mass drops sharply because the smaller grains produced by the bouncing barrier do not easily participate in the streaming instability and instead they accrete onto the star. Model \texttt{IB} also leaves more dust after the gas disk has completely dispersed inside 3 au than Model \texttt{I} (still less than $10^{-3} M_\oplus$). When the disk is about to disappear ($\sim$3 Myr), the dust mass inside 2 au is completely dominated by sub-micron grains. We note that collisions between planetesimals may generate an additional (debris) dust component which we do not consider in these evolution models.

%-----------------------------------------------------------------------------%
% Midplane turbulence
%-----------------------------------------------------------------------------%
\subsection{Midplane turbulence}
\label{sec:results:IA}

All the models presented so far have the turbulence parameter fixed at $\av = \at = 10^{-2}$. As we discussed in section \ref{sec:methods:IA}, turbulence in the midplane may be better represented by low values of $\at = 10^{-4}$. So in Model \texttt{IA} (Table \ref{tab:params}) we use two parameters: $\av = 10^{-2}$ to compute the viscous evolution of the disk, and $\at = 10^{-4}$ to compute the collision speeds between solid particles. The fragmentation-limited Stokes number scales as $1 / \at$ (Equation (\ref{eqn:Stfrag})). In Model \texttt{IAB} the grain sizes are much larger than in Model \texttt{I}, but note that in \texttt{IA} the particle growth is drift-limited beyond $\sim$ 1 au.

Naively one might expect that larger grains would lead to more planetesimal formation, but that does not seem to be the case. Model \texttt{IA} actually produces the fewest planetesimals (Table \ref{tab:results}). This is because of the effects of larger grains on the mass loss rate  due to FUV photoevaporation. As discussed in \citet{Gorti_2015}, changing the cross-sectional grain area leads to two effects: (i) a reduction in FUV attenuation and deeper penetration of photons resulting in  a higher flow density, and (ii) reduction in the FUV heating due to small grains resulting in lower flow temperatures. These oppose each other ($\dot{\Sigma}_{\rm pe}
\propto \rhog \sqrt{T}$) and hence the photoevaporation rate only depends weakly on grain size with the temperature effect being marginally stronger. Larger grain sizes therefore result in slightly lower mass loss rates, and hence longer disk survival times (also see Figure 4 of \citet{Gorti_2015}). This is seen in 
Figure \ref{fig:killer-SI-IA} where the reduction in the FUV photoevaporation rate is seen to prolong the disk lifetime by 0.82 Myr (based on the time when 99.9\% of the disk gas has disappeared). A long disk lifetime is not inherently detrimental to planetesimal formation, but with less FUV photoevaporation, the dust-to-gas ratio cannot easily reach the value required by the streaming instability across most of the disk.

%-----------------------------------------------------------------------------%
% Complete model
%-----------------------------------------------------------------------------%
\subsection{Complete model}
\label{sec:results:IAB}

Finally, Model \texttt{IAB} (Table \ref{tab:params}) includes the water ice line, the bouncing barrier of Model \texttt{IB}, and the reduced midplane turbulence of Model \texttt{IA}. This is our most complete model. Its most salient feature is that it actually produces a large population of planetesimals in the terrestrial zone. While other models can barely reach $0.4 M_\oplus$ inside 3 au, Model \texttt{IAB} produces $8.4 M_\oplus$ -- easily enough to reproduce the terrestrial planets in the solar system (Table \ref{tab:results}). In this section we investigate how the combination of the bouncing barrier and low midplane turbulence can increase the planetesimal mass inside the terrestrial zone by an order of magnitude compared to Model \texttt{I}.

\begin{figure*}
	\centering
	\includegraphics[width=1.9\columnwidth]{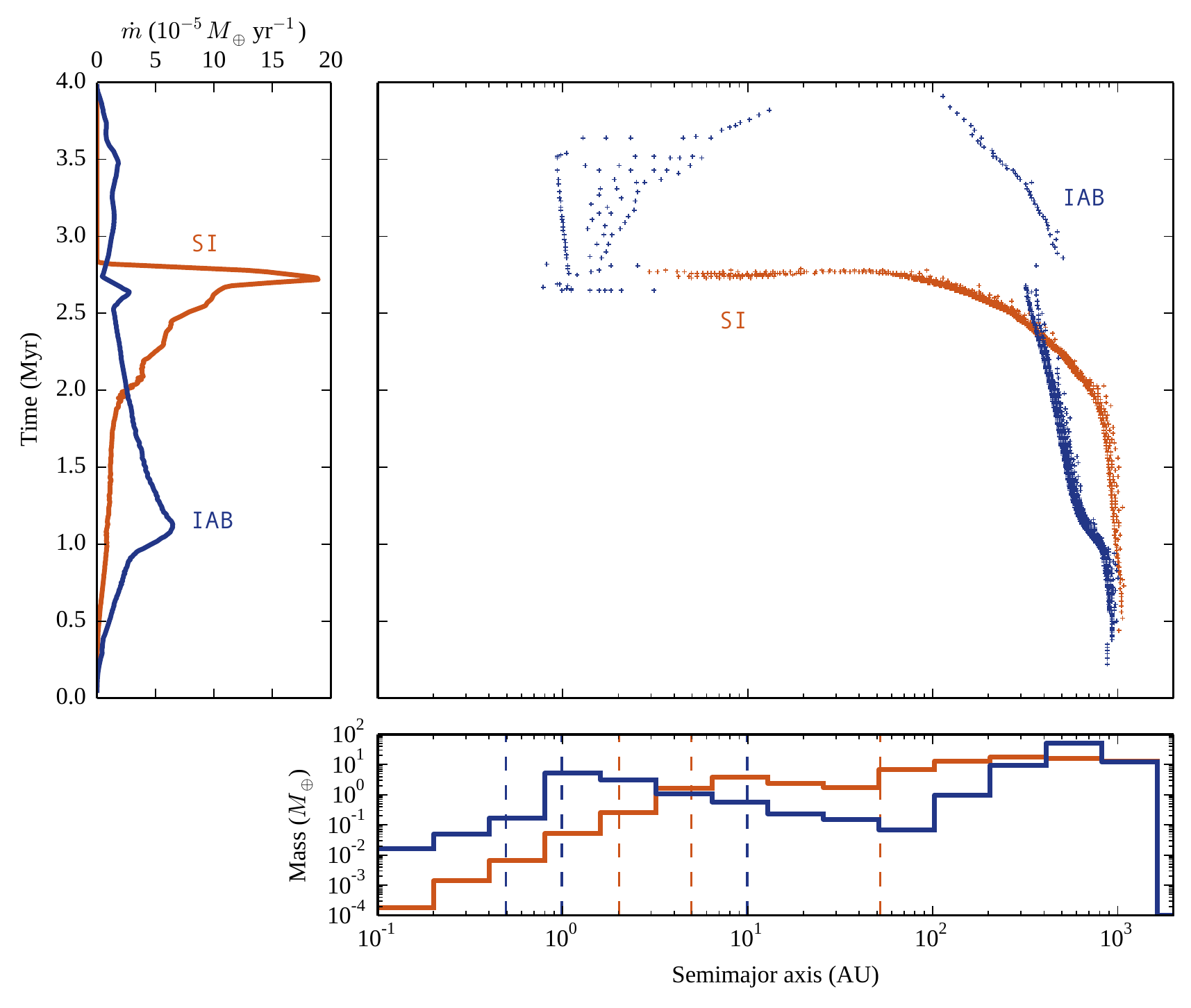}
    \caption{Time and place where planetesimals form in models \texttt{SI} (orange) and \texttt{IAB} (dark blue). Model \texttt{IAB} is our most complete model. It includes both the reduced midplane turbulence of Model \texttt{IA} and the bouncing barrier of Model \texttt{IB}. The bouncing barrier results in an increase in small grains in the inner disk and therefore an increased FUV photoevaporation rate. In Model \texttt{IAB}, an inner cavity begins to form while much of the disk is still present. With a decrease in accretion, the sudden drop in FUV photons halts planetesimal formation in the outer disk. Over the next few hundred thousand years, the disk expands viscously until FUV photoevaporation becomes effective again. In the inner disk, 8.4 $M_\oplus$ of planetesimals form in the inner 3 au.}
    \label{fig:killer-SI-IAB}
\end{figure*}

We begin with a look at the spacetime diagram of Model \texttt{IAB}, shown in Figure \ref{fig:killer-SI-IAB}. For the first few million years, the results look nearly identical to Model \texttt{IA} (shown in Figure \ref{fig:killer-SI-IA}). However, the reduced midplane turbulence in this case also results in larger grains and slightly longer disk lifetimes. Larger grains also result in more drift and increase the solids-to-gas mass fraction slightly. The addition of the bouncing barrier in the inner disk makes grains smaller in Model \texttt{IAB} and mitigates the loss of dust due to radial drift. The smaller grains also result in higher FUV photoevaporation rates in the inner disk compared to the other models. All these effects combine to produce a different evolutionary sequence in Model \texttt{IAB}. After 2.7 Myr the outer disk suddenly stops producing planetesimals for 60,000 years. Then planetesimals start to form again at a slower rate, and farther out. This break occurs just as the inner disk starts forming planetesimals. In the inner disk, planetesimal formation follows a different pattern from previous models. Planetesimals begin to form inside a narrow band that gradually expands both inward and outward.

\begin{figure*}
	\centering
	\includegraphics[width=1.9\columnwidth]{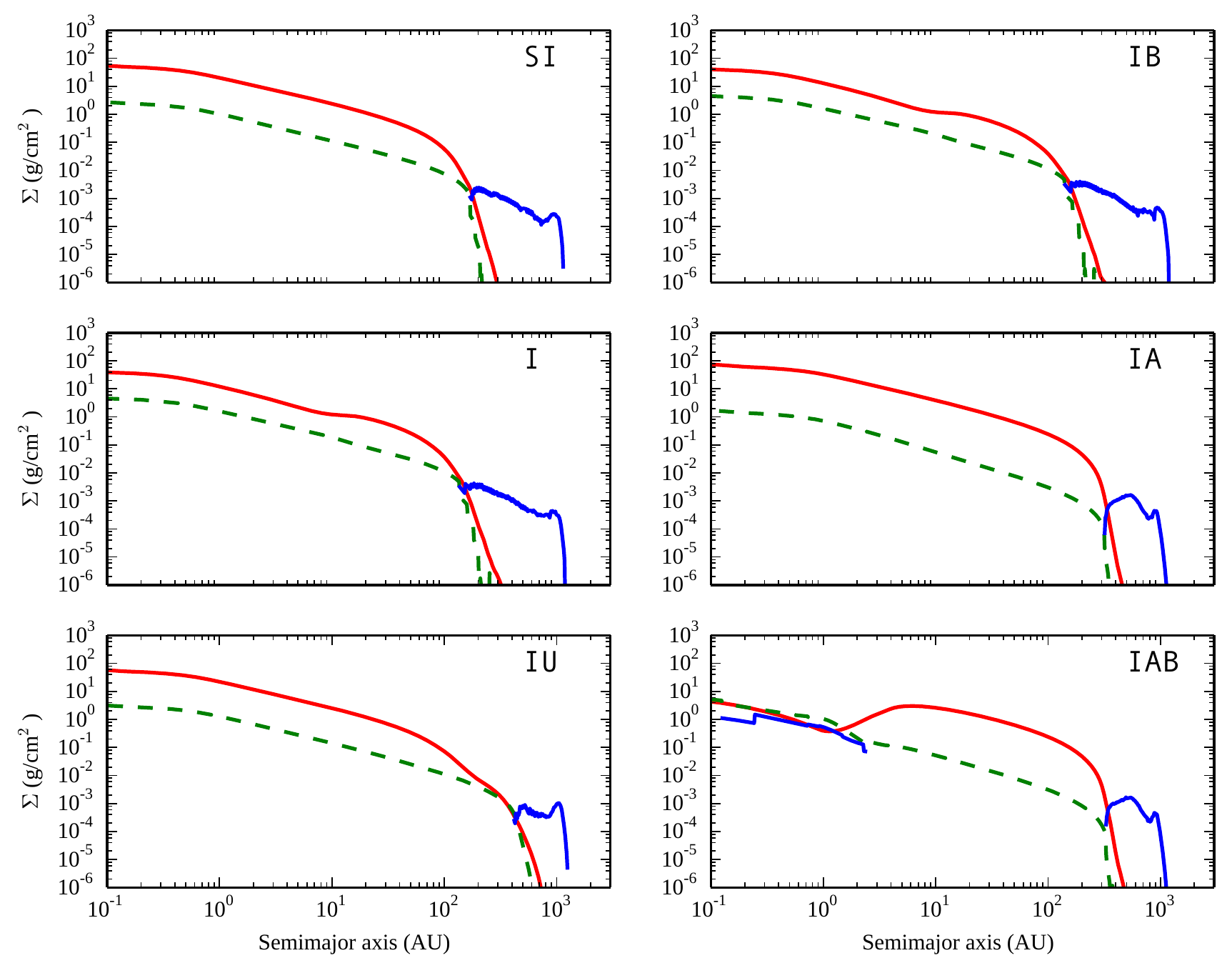}
    \caption{A snapshot of the surface density profile for the gas (solid red), dust (dashed green), and planetesimals (solid blue) components of all the runs, taken at 2.6 Myr. Model \texttt{IAB} forms an inner cavity earlier than the other models. The inner cavity prevents further loss of solid material through accretion.}
    \label{fig:cavity-all}
\end{figure*}

Figure \ref{fig:cavity-all} shows what is happening in Model \texttt{IAB} 100,000 years before the sudden change in the disk. All the other runs are also included for comparison. The figure shows the surface density profile of the gas, dust, and planetesimal components of each model. What sets Model \texttt{IAB} apart from the others is that it starts forming an inner cavity already at 2.6 Myr, while there are still 17 $M_\oplus$ of dust remaining in the disk.  As the disk evolves, the bouncing barrier produces small grains and increases FUV photoevaporation in the inner disk. A gap begins to form where the photoevaporation rate exceeds the accretion rate. This decreases the gas surface density in the inner disk. When the inner disk depletes, accretion onto the star decreases sharply (since $\dot{M}_{\rm acc}\propto \Sigmag$). This has two key consequences:

\begin{itemize}
\item When the accretion rate decreases, the accretion-generated FUV flux also drops dramatically. This effectively shuts down FUV photo-evaporation in the outer disk. When that happens, the disk begins to expand viscously. At some point, the outer edge becomes distant enough that the lower FUV emission can once again unbind the gas molecules. This causes the hiatus in planetesimal formation that we observe in Figure \ref{fig:killer-SI-IAB}.

\item In the inner disk, loss of gas raises the dust-to-gas ratio. Photoevaporation by FUV and X-rays continues and the solid grains eventually reach a dust-to-gas ratio and Stokes number that allow the streaming instability to form planetesimals. Admittedly, the streaming instability has not been studied well in environments with extreme dust-to-gas ratios, but if the streaming instability did not occur, the dust would form an ever denser, ever thinner disk that would eventually become gravitationally unstable in some Toomre-like instability.
\end{itemize}

Figure \ref{fig:killer-SI-IAB} also gives us a clue of how the final stages of the disk evolution occur. The planetesimals in the spacetime diagram very roughly trace the edges of the disk. After 2.6 Myr, the inner cavity forms and starts to expand outward. After 2.7 Myr FUV photo-evaporation is re-ignited in the outer disk, and the outer edge once gain starts to move inward. As the two disk edges evolve, they leave behind a trail of planetesimals. After 4 Myr the two fronts meet, and the disk is fully dissipated.

\begin{figure}
	\centering
	\includegraphics[width=\columnwidth]{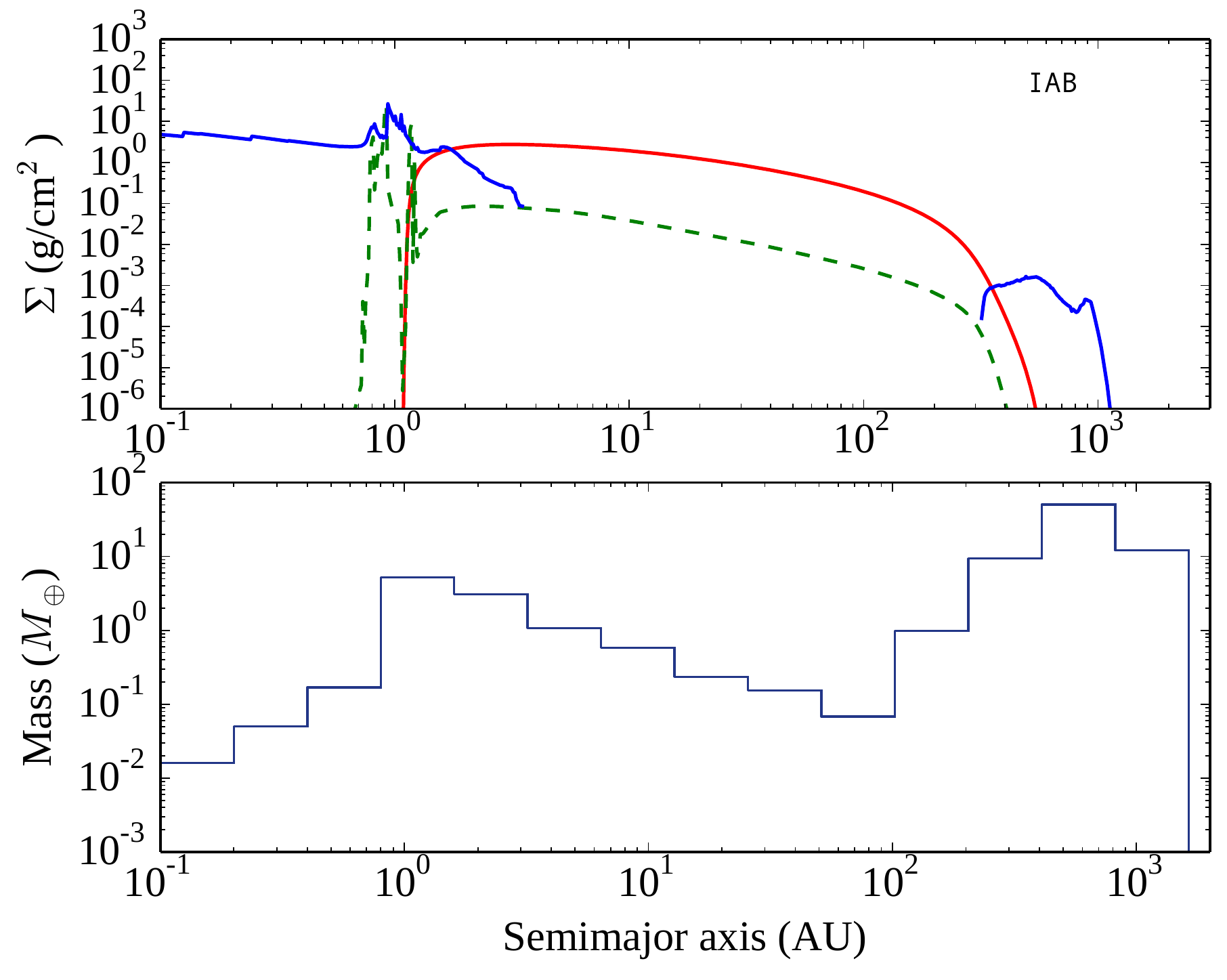}
    \caption{The top plot shows the surface density profile of the gas (solid red), dust (dashed green), and planetesimals (solid blue) components of Model \texttt{IAB}. The snapshot is taken at 2.7 Myr. At this time, the gap is fully formed, but has expanded. The bottom plot shows the total planetesimal mass in Model \texttt{IAB} \textit{at the end of the run}. The figure shows that the planetesimals in the terrestrial zone mostly form just outside the cavity.}
    \label{fig:cavity-IAB}
\end{figure}

Figure \ref{fig:cavity-IAB} helps complete the picture in the inner disk. The top plot of the figure shows the gas density profile in Model \texttt{IAB} after 2.7 Myr --- just at the moment when the inner cavity has fully formed. The bottom plot shows the mass distribution of planetesimals at the end of the simulation. This figure makes it clear that the planetesimals form just outside the gap. This is what we would expect if the underlying mechanism is that the gap prevents the accretion of solids onto the star.

Lastly, what makes Model \texttt{IAB} produce an early gap while the other models do not? The answer seems to be that the bouncing barrier promotes FUV photoevaporation in the inner disk (by creating very small grains) while in the outer disk, the larger particles inhibit FUV photoevaporation and cause the disk to last longer.

%%%%%%%%%%%%%%%%%%%%%%%%%%%%%%%%%%%%%%%%%%%%%%%%%%%%%%%%%%%%%%%%%%%%%%%%%%%%%%%
%
% OTHER ISSUES
%
%%%%%%%%%%%%%%%%%%%%%%%%%%%%%%%%%%%%%%%%%%%%%%%%%%%%%%%%%%%%%%%%%%%%%%%%%%%%%%%
\section{Other factors}
\label{sec:other}

Planetesimal formation is a complex, multi-physics problem. We have shown that photoevaporation, dust evolution, the water ice line, and the streaming instability all probably play an important role in the formation of planetesimals. In this section we discuss various physical processes that we could not model adequately in our simulations, and we speculate on how these processes might change our results.

%-----------------------------------------------------------------------------%
% DISK VISCOSITY
%-----------------------------------------------------------------------------%
\subsection{Disk viscosity}
\label{sec:other:viscosity}

We note that all the results presented here are sensitive to assumptions made on viscous disk evolution. We have chosen a simple $\alpha$-parametrization of disk viscosity, and moreover assume that $\alpha$ is constant in radius and time as the disk evolves. These assumptions may not be realistic, but a more accurate prescription of disk viscosity is still not available, although considerable progress is being made in this direction \citep[e.g.][]{Bai_2016a,Simon_2015}. Disk lifetimes due to a combination of viscous evolution and photoevaporation also depend on the choice of the numerical value of $\alpha$ \citep[see][for a more in-depth discussion]{Gorti_2015}, a value of 0.01 best reproduces most observed characteristics, such as the overall disk lifetime, stellar mass accretion rate and accretion luminosity.  Viscous evolution sets the basic timescales in the disk (more so at earlier epochs when photoevaporation rates are comparatively low), and all physical processes modelled here are also subject to resultant uncertainties.

We conducted simulations with lower viscosity ($\av = 10^{-3}$). As shown in \citet{Gorti_2015}, the lower accretion luminosity and viscous spreading result in longer disk evolution times. Our key result (that most planetesimals form in the outer disk) remained unchanged. FUV photoevaporation occurs everywhere in the disk. Its presence is most easily seen in the outer disk because the gas removed is not replenished; the dust-to-gas ratio increases and planetesimal formation occurs. Elsewhere, the gas that photoevaporates is replaced by gas that viscously accretes from outside, so the dust-to-gas ratio remains low.

%-----------------------------------------------------------------------------%
% LAMINAR ACCRETION
%-----------------------------------------------------------------------------%
\subsection{Laminar accretion}
\label{sec:other:laminar}

In models \texttt{I} and \texttt{IA}, the pile-up of solids inside the ice line is frustrated by the viscous accretion that drags these solids into the star. This is an artefact of our one-dimensional disk model, which forces the entire gas column to accrete at the same rate, set by $\av = 10^{-2}$. In model \texttt{IA} we can capture the low collision speeds from the $\at = 10^{-4}$ on the midplane, but the gas component is still forced to evolve with $\av = 10^{-2}$. This means the model greatly overestimates the gas accretion rate on the midplane. Outside the ice line this is not much of an issue because the radial speed of the solids is dominated by the radial drift relative to the gas (Equation (\ref{eqn:u_drift})). But inside the ice line, where the bouncing barrier keeps solids closely coupled to the gas, our models greatly overestimate the radial speed of solids. In fact, a 2D model by \citet{Ciesla_2007} shows that there is even a region of \textit{outward} gas flow around the midplane, and an outward-transport region where the viscous flow pushes solids away from the star.

With a more realistic model that could capture the laminar structure of the disk, we would expect to see a much larger pile-up of solids close to the ice line, and a much larger planetesimal mass in the terrestrial zone.

%-----------------------------------------------------------------------------%
% ROCHE DENSITY
%-----------------------------------------------------------------------------%
\subsection{Roche density}
\label{sec:other:roche}

In this section we take a closer look at the formal conditions for particle collapse. As noted in the introduction, the streaming instability belongs to a family of models that aim to produce over-densities of solid particles to the point of initiating gravitational collapse. In the case of the streaming instability, these over-densities take the form of elongated filaments that stretch along the azimuthal direction. The formal (but approximate) condition for gravitational collapse is that the midplane density of solid particles $\rhop$ must reach the Roche density,

\begin{equation}\label{eqn:roche}
	\rhoR = \frac{9 \Omega^2}{4 \pi G} = \frac{9}{4 \pi} \frac{M_\star}{r^3},
\end{equation}
where $\Omega$ is the Keplerian frequency, $G$ is the gravitational constant, $M_\star$ is the mass of the central star, and $r$ is the orbital distance. When $\rhop > \rhoR$, the particle self-gravity overwhelms Keplerian shear, leading to collapse \citep{Goldreich_1973,Johansen_2014}. This means that, even when the streaming instability effectively forms a stable filament of particles, that filament will not produce any planetesimals unless $\rhop > \rhoR$. Since the streaming instability is largely a scale-free process, the particle density inside the filaments is proportional to the gas density at the time that the streaming instability occurs,

\begin{equation}\label{eqn:rhopSI}
	\rho_{\rm p,SI} = \epsilon \; \rhog,
\end{equation}
where $\rhog$ is the gas density at the midplane, and $\rho_{\rm p,SI}$ is the particle density produced by the streaming instability inside a filament. The value of $\epsilon$ probably depends on the Stokes number, with smaller particles forming less dense filaments \citep{Carrera_2015}. As we discussed in section \ref{sec:methods:SI}, the streaming instability can only occur when $\rhop / \rhog > 1$ on the midplane. The streaming instability then proceeds to concentrate particles into denser filaments, so that $\epsilon = \rho_{\rm p,SI} / \rhog \gg 1$ inside the filaments. However, the value of $\epsilon$ has only been measured precisely for $\St = 0.3$ particles, which have $\epsilon$ going locally up to at least $10^4$ \citep{Johansen_2015}.

As mentioned in the discussion, the streaming instability occurs whenever $\Sigmap / \Sigmag > \Zcrit$, where $\Zcrit$ is a function of the particle Stokes number. From the point of view of the streaming instability, it makes no difference if $\Zcrit$ is achieved by removing gas or by adding solids. However, Equation (\ref{eqn:rhopSI}) shows that it is easier to reach $\rho_{\rm p,SI} > \rhoR$ if the streaming instability can be triggered at a higher gas density.

To proceed with this investigation we consider a toy model in which we calculate $\rhog$, $Z$, St, and $\Zcrit(\St)$ at 2.5 Myr (i.e.~shortly before the inner disk begins to photoevaporate) and ignore any subsequent change in St or the solid density. In this toy model, St and $\rhop$ are fixed, and the only way to trigger the streaming instability is to increase $Z$ by removing gas. In other words, the gas component has to be depleted by a factor of $Z/\Zcrit$ for the streaming instability to occur,

\begin{equation}\label{eqn:rhog_crit}
	\rho_{\rm g,crit} = \rho_{\rm g,2.5 Myr} \, \frac{Z_{\rm 2.5 Myr}}{\Zcrit},
\end{equation}
where $\rho_{\rm g,crit}$ is the (critical) midplane density when the streaming instability begins. We then compute $\rho_{\rm p,SI}$ assuming $\epsilon = 10^4$,

\begin{equation}\label{eqn:rhopSI-model}
	\rho_{\rm p,SI} = 10^4 \, \rho_{\rm g,crit}
\end{equation}
Except for the value of $\epsilon$, which is not well understood for the wide range in Stokes numbers considered here, this toy model is conservative. Larger Stokes numbers imply lower $\Zcrit$ and larger $\rho_{\rm p,SI}$.

\begin{figure*}
	\centering
	\includegraphics[width=1.8\columnwidth]{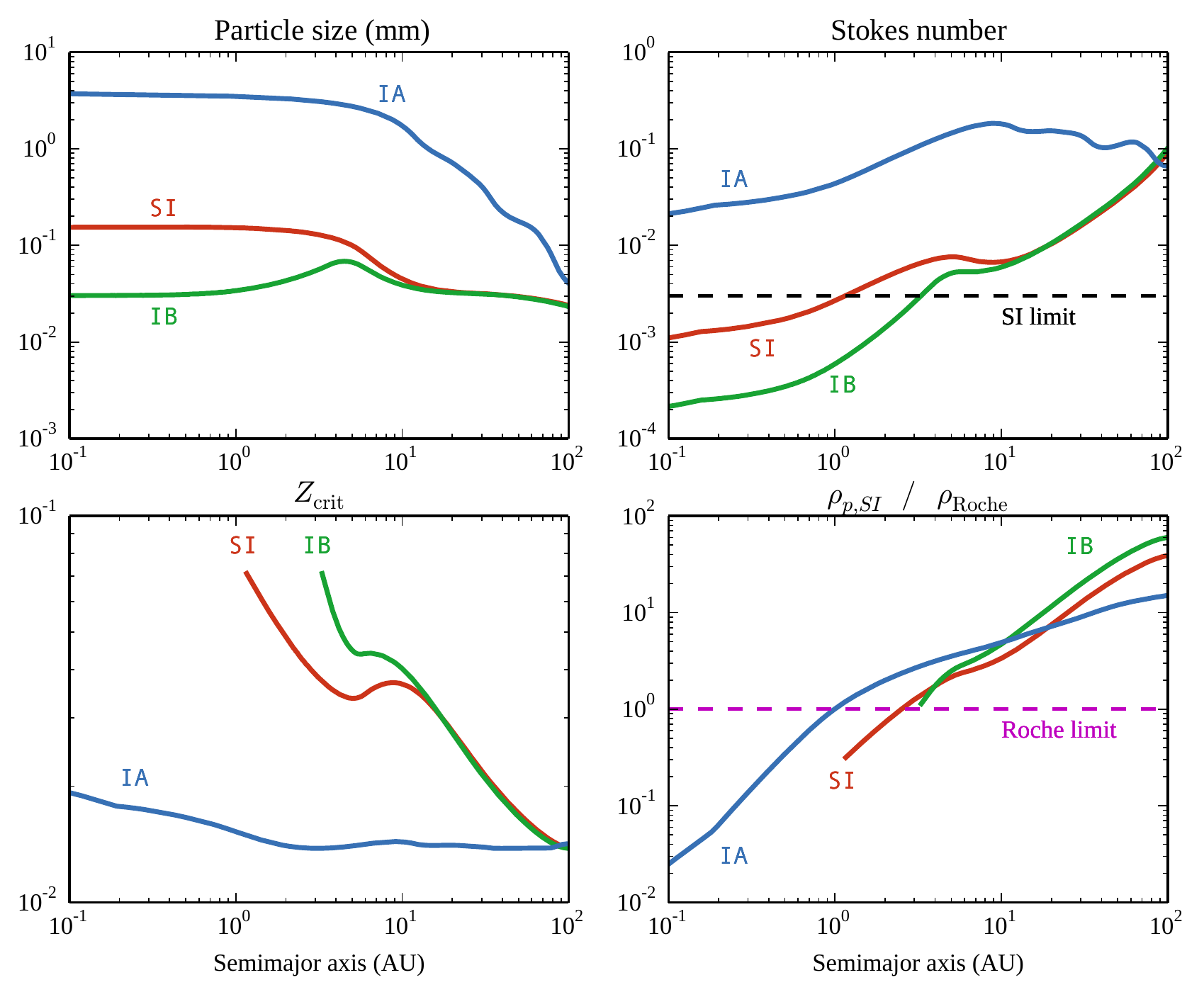}
    \caption{Snapshot of Models \texttt{SI} (red), \texttt{IB} (green), and \texttt{IA} (blue) at 2.5 Myr. This is just before photoevaporation starts to become significant inside 100 au. The top-left and top-right plots show the mass-weighted mean particle size and the corresponding Stokes number (Equation (\ref{eqn:St})). The dashed line at $\St = 0.003$ marks the smallest Stokes number that has been seen to participate in the streaming instability \citep{Carrera_2015}. In the bottom plots we only show results for $\St \ge 0.003$. The bottom-left plot shows the minimum dust-to-gas ratio $\Zcrit$ required to trigger the streaming instability at that Stokes number. The bottom-right shows the particle density produced by the streaming instability ($\rho_{\rm p,SI}$) over the Roche density in a toy model where $\St$ and $\Sigmap$ remain fixed, and $\rho_{\rm p,SI} = 10^4 \rho_{\rm g}$ as observed by \citet{Johansen_2015}. In a real disk there are many factors that may increase or decrease $\rho_{\rm p,SI}$ (see main text).}
    \label{fig:roche}
\end{figure*}

The implications of this toy model are shown in Figure \ref{fig:roche}. The figure shows a snapshot of three sample runs at 2.5 Myr. The plot on the top-left shows the mass-weighted mean particle size,

\begin{equation}
	\langle a \rangle = \sum_{i=1}^N \frac{ a_i \Sigma_{{\rm p},i} }{\Sigmap},
\end{equation}
where $a_{\rm i}$ and $\Sigma_{{\rm p},i}$ are the size and surface density of particles in the $i^{\rm th}$ particle bin, and $\Sigmap$ is the total particle surface density. The plot on the top-right shows the Stokes number for particles of size $\langle a \rangle$. The dashed line marks the limit $\St = 0.003$, which is the smallest particle size that has been observed to participate in the streaming instability \citep{Carrera_2015}. This plot seems to imply that models \texttt{SI} and \texttt{IB} cannot form particle filaments inside 3 au. We remind the reader that the onset of photoevaporation in the inner disk leads to an increase in the particle Stokes numbers inside 3 au. We chose to exclude this fact in this section to keep the toy model simple. In the bottom two plots we only show results in places where $\St \ge 0.003$ before photoevaporation.

The bottom-left plot of Figure \ref{fig:roche} shows the critical dust-to-gas ratio $\Zcrit$ assuming a uniform particle size $\langle a \rangle$. Finally, we show $\rho_{\rm p,SI} / \rhoR$ in the bottom-right plot, following Equations (\ref{eqn:roche}) and (\ref{eqn:rhopSI-model}). We also mark the critical value $\rho_{\rm p,SI} / \rhoR = 1$, above which the filaments produced by the streaming instability should collapse into planetesimals. The plot shows that model \texttt{IA}, benefiting from large particles, can easily form planetesimals well inside the terrestrial zone. The other models cannot form planetesimals unless the Stokes numbers increase.

One feature of the bouncing barrier that we did not implement is that it is a ``soft'' barrier -- when gas dissipates, the particles do not fragment into smaller ones, so their $\St$ increases. As $\rhog \rightarrow 0$, $\St$ approaches the fragmentation barrier. To conclude this section, we list three more processes that may significantly increase $\rhop$ inside a filament and have not been adequately explored in the literature.

\begin{itemize}
\item Inside the filament, orbits are strongly Keplerian and collision speeds between particles may be much smaller than $\cs \sqrt{\alpha \St}$ (Equation (\ref{eqn:du_coll})). If so, the Stokes number will no longer be fixed by Equation (\ref{eqn:Stfrag}), and coagulation will proceed to larger Stokes numbers. As particles grow, collision speeds would continue to drop. See \citet{Carrera_2015} for an earlier exploration of this concept.

\item Even when a stable filament fails to reach the Roche density, $\rhop$ will continue to grow on the radial drift timescale as solid particles in the outer disk migrate inward and become trapped when they enter the filament. This process will only work if the drift timescale is shorter than the evaporation timescale, which is typically the case.

\item Finally, we do not know how a filament produced by the streaming instability will respond to the removal of gas by photoevaporation. At some point the gas component should not be able to support the full vertical height of the filament and $\rhop$ should begin to increase. As $\rhog \rightarrow 0$, $\rhop \rightarrow \infty$ on the midplane and the filament may experience a 1-dimensional instability along the lines of \citet{Goldreich_1973}, \citet{Sekiya_1998}, and \citet{Youdin_2002}.
\end{itemize}

The last two mechanisms in particular suggest that, even if $\rho_{\rm p,SI} < \rhoR$ initially, $\rhop$ could grow arbitrarily high and eventually reach the Roche density. In other words, once the streaming instability forms a stable filament, planetesimals may be unavoidable. The formation of filaments of low-St pebbles and the evolution of such filaments in response to gas removal are topics that should be explored more in future simulations.

%%%%%%%%%%%%%%%%%%%%%%%%%%%%%%%%%%%%%%%%%%%%%%%%%%%%%%%%%%%%%%%%%%%%%%%%%%%%%%%
%
% SUMMARY AND CONCLUSIONS
%
%%%%%%%%%%%%%%%%%%%%%%%%%%%%%%%%%%%%%%%%%%%%%%%%%%%%%%%%%%%%%%%%%%%%%%%%%%%%%%%
\section{Summary and conclusions}
\label{sec:conclusions}

In this paper we have presented a global disk evolution models that incorporate photoevaporation and the formation of planetesimals by the streaming instability. Our key result is that planetesimals are a natural by-product of protoplanetary disk evolution. Planetesimal formation does not require a fundamental change in how we understand disk evolution --- it merely requires a multi-physics view of planet formation that combines the work done in many areas of protoplanetary disk models and planetesimal formation models over recent years.  Concretely, we offer two key findings:

\begin{itemize}
\item Planetesimals form early in the outer disk as a result of the streaming instability and FUV photo-evaporation. This outcome seems unavoidable, as even the most conservative models produce at least $\sim 60 M_\oplus$ of planetesimals beyond 100 au. This result is very robust and does not require fine-tuning of any disk parameter. If these planetesimals further assimilate to form planets in the outer disk, their dynamical effects on the protoplanetary disk may be observable with high-sensitivity facilities such as ALMA.

\item The formation of planetesimals in the terrestrial zone depends mostly on the formation of an inner cavity in the disk. Before the cavity forms, the streaming instability is ineffective and cannot easily produce planetesimals. Once a cavity forms, the solid material can no longer accrete onto the star. As photoevaporation removes the remaining gas, the dust-to-gas ratio continues to increase and the remaining solids are forced to eventually form planetesimals. Our most complete model, Model \texttt{IAB} (which includes the water ice line, bouncing barrier, and low midplane turbulence), produces 8.4 $M_\oplus$ of planetesimals in the terrestrial zone. All other models produce at most 0.3 $M_\oplus$. What sets Model \texttt{IAB} apart is that it forms a cavity sooner than the others.
\end{itemize}

All our models produce very few planetesimals in the giant planet zone (3-30 au), and those that do form appear too late and could probably not grow into giant planet cores before gas dissipation \citep{Bitsch_2015}. It appears that the seed planetesimals that form giant planets must form through some other type of process, perhaps related to early particle pile-ups near ice lines \citep{Sirono_2011a,Sirono_2011b}.

Finally, our results suggest new questions and new opportunities for future research:

\begin{itemize}
\item How do disk winds affect planetesimal formation by the streaming instability?

\item How does a particle filament produced by the streaming instability evolve if it does not initially reach the Roche density?

\item What is the collision speed between particles inside those filaments?
\end{itemize}

%%%%%%%%%%%%%%%%%%%%%%%%%%%%%%%%%%%%%%%%%%%%%%%%%%%%%%%%%%%%%%%%%%%%%%%%%%%%%%%
%
% ACKNOWLEDGEMENTS
%
%%%%%%%%%%%%%%%%%%%%%%%%%%%%%%%%%%%%%%%%%%%%%%%%%%%%%%%%%%%%%%%%%%%%%%%%%%%%%%%
\section*{Acknowledgements}

D.C. acknowledges Chao-Chin Yang for many helpful discussions and for his insights on the Roche density and the streaming instability. The authors acknowledge the support from the Knut and Alice Wallenberg Foundation (grants 2012.0150, 2014.0017, 2014.0048), the Swedish Research Council (grants 2011-3991 and 2014-5775), and the European Research Council Starting Grant 278675-PEBBLE2PLANET. U.G. is supported by the National Aeronautics and Space Administration through the NASA Astrobiology Institute under Co-operation Agreement Notice NNH13ZDA017C issued through the Science Mission Directorate, and a grant from the National Science Foundation (AST1313003). This project made use of NASA HEC supercomputing resources.

%%%%%%%%%%%%%%%%%%%% REFERENCES %%%%%%%%%%%%%%%%%%

\bibliographystyle{aasjournal}
\bibliography{references}

\begin{thebibliography}{}
\expandafter\ifx\csname natexlab\endcsname\relax\def\natexlab#1{#1}\fi

\bibitem[{{Adams} {et~al.}(2008){Adams}, {Seager}, \&
  {Elkins-Tanton}}]{Adams_2008}
{Adams}, E.~R., {Seager}, S., \& {Elkins-Tanton}, L. 2008, \apj, 673, 1160

\bibitem[{{Adams} {et~al.}(2004){Adams}, {Hollenbach}, {Laughlin}, \&
  {Gorti}}]{Adams_2004}
{Adams}, F.~C., {Hollenbach}, D., {Laughlin}, G., \& {Gorti}, U. 2004, \apj,
  611, 360

\bibitem[{{Alexander} {et~al.}(2014){Alexander}, {Pascucci}, {Andrews},
  {Armitage}, \& {Cieza}}]{Alexander_2014}
{Alexander}, R., {Pascucci}, I., {Andrews}, S., {Armitage}, P., \& {Cieza}, L.
  2014, Protostars and Planets VI, 475

\bibitem[{{Alexander} \& {Armitage}(2009)}]{Alexander_2009}
{Alexander}, R.~D., \& {Armitage}, P.~J. 2009, \apj, 704, 989

\bibitem[{{Alexander} {et~al.}(2006{\natexlab{a}}){Alexander}, {Clarke}, \&
  {Pringle}}]{Alexander_2006a}
{Alexander}, R.~D., {Clarke}, C.~J., \& {Pringle}, J.~E. 2006{\natexlab{a}},
  \mnras, 369, 216

\bibitem[{{Alexander} {et~al.}(2006{\natexlab{b}}){Alexander}, {Clarke}, \&
  {Pringle}}]{Alexander_2006b}
---. 2006{\natexlab{b}}, \mnras, 369, 229

\bibitem[{{Andrews} \& {Williams}(2005)}]{Andrews_2005}
{Andrews}, S.~M., \& {Williams}, J.~P. 2005, \apj, 631, 1134

\bibitem[{{Bai}(2016)}]{Bai_2016b}
{Bai}, X.-N. 2016, \apj, 821, 80

\bibitem[{{Bai} \& {Stone}(2010)}]{Bai_2010}
{Bai}, X.-N., \& {Stone}, J.~M. 2010, \apj, 722, 1437

\bibitem[{{Bai} \& {Stone}(2013)}]{Bai_2013}
---. 2013, \apj, 769, 76

\bibitem[{{Bai} {et~al.}(2016){Bai}, {Ye}, {Goodman}, \& {Yuan}}]{Bai_2016a}
{Bai}, X.-N., {Ye}, J., {Goodman}, J., \& {Yuan}, F. 2016, \apj, 818, 152

\bibitem[{{Barge} \& {Sommeria}(1995)}]{Barge_1995}
{Barge}, P., \& {Sommeria}, J. 1995, \aap, 295, L1

\bibitem[{{Batalha} {et~al.}(2013){Batalha}, {Rowe}, {Bryson}, {Barclay},
  {Burke}, {Caldwell}, {Christiansen}, {Mullally}, {Thompson}, {Brown},
  {Dupree}, {Fabrycky}, {Ford}, {Fortney}, {Gilliland}, {Isaacson}, {Latham},
  {Marcy}, {Quinn}, {Ragozzine}, {Shporer}, {Borucki}, {Ciardi}, {Gautier},
  {Haas}, {Jenkins}, {Koch}, {Lissauer}, {Rapin}, {Basri}, {Boss}, {Buchhave},
  {Carter}, {Charbonneau}, {Christensen-Dalsgaard}, {Clarke}, {Cochran},
  {Demory}, {Desert}, {Devore}, {Doyle}, {Esquerdo}, {Everett}, {Fressin},
  {Geary}, {Girouard}, {Gould}, {Hall}, {Holman}, {Howard}, {Howell},
  {Ibrahim}, {Kinemuchi}, {Kjeldsen}, {Klaus}, {Li}, {Lucas}, {Meibom},
  {Morris}, {Pr{\v s}a}, {Quintana}, {Sanderfer}, {Sasselov}, {Seader},
  {Smith}, {Steffen}, {Still}, {Stumpe}, {Tarter}, {Tenenbaum}, {Torres},
  {Twicken}, {Uddin}, {Van Cleve}, {Walkowicz}, \& {Welsh}}]{Batalha_2013}
{Batalha}, N.~M., {Rowe}, J.~F., {Bryson}, S.~T., {et~al.} 2013, \apjs, 204, 24

\bibitem[{{Bell} {et~al.}(2013){Bell}, {Naylor}, {Mayne}, {Jeffries}, \&
  {Littlefair}}]{Bell_2013}
{Bell}, C.~P.~M., {Naylor}, T., {Mayne}, N.~J., {Jeffries}, R.~D., \&
  {Littlefair}, S.~P. 2013, \mnras, 434, 806

\bibitem[{{Birnstiel} {et~al.}(2009){Birnstiel}, {Dullemond}, \&
  {Brauer}}]{Birnstiel_2009}
{Birnstiel}, T., {Dullemond}, C.~P., \& {Brauer}, F. 2009, \aap, 503, L5

\bibitem[{{Birnstiel} {et~al.}(2012){Birnstiel}, {Klahr}, \&
  {Ercolano}}]{Birnstiel_2012}
{Birnstiel}, T., {Klahr}, H., \& {Ercolano}, B. 2012, \aap, 539, A148

\bibitem[{{Birnstiel} {et~al.}(2011){Birnstiel}, {Ormel}, \&
  {Dullemond}}]{Birnstiel_2011}
{Birnstiel}, T., {Ormel}, C.~W., \& {Dullemond}, C.~P. 2011, \aap, 525, A11

\bibitem[{{Bitsch} {et~al.}(2015){Bitsch}, {Lambrechts}, \&
  {Johansen}}]{Bitsch_2015}
{Bitsch}, B., {Lambrechts}, M., \& {Johansen}, A. 2015, \aap, 582, A112

\bibitem[{{Blandford} \& {Payne}(1982)}]{Blandford_1982}
{Blandford}, R.~D., \& {Payne}, D.~G. 1982, \mnras, 199, 883

\bibitem[{{Borucki} {et~al.}(2011){Borucki}, {Koch}, {Basri}, {Batalha},
  {Brown}, {Bryson}, {Caldwell}, {Christensen-Dalsgaard}, {Cochran}, {DeVore},
  {Dunham}, {Gautier}, {Geary}, {Gilliland}, {Gould}, {Howell}, {Jenkins},
  {Latham}, {Lissauer}, {Marcy}, {Rowe}, {Sasselov}, {Boss}, {Charbonneau},
  {Ciardi}, {Doyle}, {Dupree}, {Ford}, {Fortney}, {Holman}, {Seager},
  {Steffen}, {Tarter}, {Welsh}, {Allen}, {Buchhave}, {Christiansen}, {Clarke},
  {Das}, {D{\'e}sert}, {Endl}, {Fabrycky}, {Fressin}, {Haas}, {Horch},
  {Howard}, {Isaacson}, {Kjeldsen}, {Kolodziejczak}, {Kulesa}, {Li}, {Lucas},
  {Machalek}, {McCarthy}, {MacQueen}, {Meibom}, {Miquel}, {Prsa}, {Quinn},
  {Quintana}, {Ragozzine}, {Sherry}, {Shporer}, {Tenenbaum}, {Torres},
  {Twicken}, {Van Cleve}, {Walkowicz}, {Witteborn}, \& {Still}}]{Borucki_2011}
{Borucki}, W.~J., {Koch}, D.~G., {Basri}, G., {et~al.} 2011, \apj, 736, 19

\bibitem[{{Brauer} {et~al.}(2008){Brauer}, {Dullemond}, \&
  {Henning}}]{Brauer_2008}
{Brauer}, F., {Dullemond}, C.~P., \& {Henning}, T. 2008, \aap, 480, 859

\bibitem[{{Calvet} \& {Gullbring}(1998)}]{Calvet_1998}
{Calvet}, N., \& {Gullbring}, E. 1998, \apj, 509, 802

\bibitem[{{Carrera} {et~al.}(2015){Carrera}, {Johansen}, \&
  {Davies}}]{Carrera_2015}
{Carrera}, D., {Johansen}, A., \& {Davies}, M.~B. 2015, \aap, 579, A43

\bibitem[{{Chambers} \& {Wetherill}(1998)}]{Chambers_1998}
{Chambers}, J.~E., \& {Wetherill}, G.~W. 1998, \icarus, 136, 304

\bibitem[{{Ciesla}(2007)}]{Ciesla_2007}
{Ciesla}, F.~J. 2007, Science, 318, 613

\bibitem[{{Drazkowska} {et~al.}(2016){Drazkowska}, {Alibert}, \&
  {Moore}}]{Drazkowska_2016}
{Drazkowska}, J., {Alibert}, Y., \& {Moore}, B. 2016, ArXiv e-prints,
  arXiv:1607.05734

\bibitem[{{Dunham} {et~al.}(2014){Dunham}, {Stutz}, {Allen}, {Evans},
  {Fischer}, {Megeath}, {Myers}, {Offner}, {Poteet}, {Tobin}, \&
  {Vorobyov}}]{Dunham_2014}
{Dunham}, M.~M., {Stutz}, A.~M., {Allen}, L.~E., {et~al.} 2014, Protostars and
  Planets VI, 195

\bibitem[{{Flaccomio} {et~al.}(2003){Flaccomio}, {Damiani}, {Micela},
  {Sciortino}, {Harnden}, {Murray}, \& {Wolk}}]{Falccomio_2003}
{Flaccomio}, E., {Damiani}, F., {Micela}, G., {et~al.} 2003, \apj, 582, 398

\bibitem[{{Flaherty} {et~al.}(2015){Flaherty}, {Hughes}, {Rosenfeld},
  {Andrews}, {Chiang}, {Simon}, {Kerzner}, \& {Wilner}}]{Flaherty_2015}
{Flaherty}, K.~M., {Hughes}, A.~M., {Rosenfeld}, K.~A., {et~al.} 2015, \apj,
  813, 99

\bibitem[{{Fromang} {et~al.}(2002){Fromang}, {Terquem}, \&
  {Balbus}}]{Fromang_2002}
{Fromang}, S., {Terquem}, C., \& {Balbus}, S.~A. 2002, \mnras, 329, 18

\bibitem[{{Gammie}(1996)}]{Gammie_1996}
{Gammie}, C.~F. 1996, \apj, 457, 355

\bibitem[{{Goldreich} \& {Ward}(1973)}]{Goldreich_1973}
{Goldreich}, P., \& {Ward}, W.~R. 1973, \apj, 183, 1051

\bibitem[{{Gorti} {et~al.}(2009){Gorti}, {Dullemond}, \&
  {Hollenbach}}]{Gorti_2009b}
{Gorti}, U., {Dullemond}, C.~P., \& {Hollenbach}, D. 2009, \apj, 705, 1237

\bibitem[{{Gorti} {et~al.}(2015){Gorti}, {Hollenbach}, \&
  {Dullemond}}]{Gorti_2015}
{Gorti}, U., {Hollenbach}, D., \& {Dullemond}, C.~P. 2015, \apj, 804, 29

\bibitem[{{G{\"u}del} {et~al.}(2007){G{\"u}del}, {Telleschi}, {Audard},
  {Skinner}, {Briggs}, {Palla}, \& {Dougados}}]{Gudel_2007}
{G{\"u}del}, M., {Telleschi}, A., {Audard}, M., {et~al.} 2007, \aap, 468, 515

\bibitem[{{Gullbring} {et~al.}(2000){Gullbring}, {Calvet}, {Muzerolle}, \&
  {Hartmann}}]{Gullbring_2000}
{Gullbring}, E., {Calvet}, N., {Muzerolle}, J., \& {Hartmann}, L. 2000, \apj,
  544, 927

\bibitem[{{Gundlach} \& {Blum}(2015)}]{Gundlach_2015}
{Gundlach}, B., \& {Blum}, J. 2015, \apj, 798, 34

\bibitem[{{G{\"u}ttler} {et~al.}(2010){G{\"u}ttler}, {Blum}, {Zsom}, {Ormel},
  \& {Dullemond}}]{Guttler_2010}
{G{\"u}ttler}, C., {Blum}, J., {Zsom}, A., {Ormel}, C.~W., \& {Dullemond},
  C.~P. 2010, \aap, 513, A56

\bibitem[{{Haisch} {et~al.}(2001){Haisch}, {Lada}, \& {Lada}}]{Haisch_2001}
{Haisch}, Jr., K.~E., {Lada}, E.~A., \& {Lada}, C.~J. 2001, \apjl, 553, L153

\bibitem[{{Hartmann} {et~al.}(1998){Hartmann}, {Calvet}, {Gullbring}, \&
  {D'Alessio}}]{Hartmann_1998}
{Hartmann}, L., {Calvet}, N., {Gullbring}, E., \& {D'Alessio}, P. 1998, \apj,
  495, 385

\bibitem[{{Howard} {et~al.}(2010){Howard}, {Marcy}, {Johnson}, {Fischer},
  {Wright}, {Isaacson}, {Valenti}, {Anderson}, {Lin}, \& {Ida}}]{Howard_2010}
{Howard}, A.~W., {Marcy}, G.~W., {Johnson}, J.~A., {et~al.} 2010, Science, 330,
  653

\bibitem[{{Ingleby} {et~al.}(2011){Ingleby}, {Calvet}, {Hern{\'a}ndez},
  {Brice{\~n}o}, {Espaillat}, {Miller}, {Bergin}, \& {Hartmann}}]{Ingebly_2011}
{Ingleby}, L., {Calvet}, N., {Hern{\'a}ndez}, J., {et~al.} 2011, \aj, 141, 127

\bibitem[{{Johansen} {et~al.}(2014){Johansen}, {Blum}, {Tanaka}, {Ormel},
  {Bizzarro}, \& {Rickman}}]{Johansen_2014}
{Johansen}, A., {Blum}, J., {Tanaka}, H., {et~al.} 2014, Protostars and Planets
  VI, 547

\bibitem[{{Johansen} \& {Klahr}(2005)}]{Johansen_2005}
{Johansen}, A., \& {Klahr}, H. 2005, \apj, 634, 1353

\bibitem[{{Johansen} {et~al.}(2011){Johansen}, {Klahr}, \&
  {Henning}}]{Johansen_2011}
{Johansen}, A., {Klahr}, H., \& {Henning}, T. 2011, \aap, 529, A62

\bibitem[{{Johansen} {et~al.}(2015){Johansen}, {Mac Low}, {Lacerda}, \&
  {Bizzarro}}]{Johansen_2015}
{Johansen}, A., {Mac Low}, M.-M., {Lacerda}, P., \& {Bizzarro}, M. 2015,
  Science Advances, 1, 1500109

\bibitem[{{Johansen} {et~al.}(2007){Johansen}, {Oishi}, {Mac Low}, {Klahr},
  {Henning}, \& {Youdin}}]{Johansen_2007}
{Johansen}, A., {Oishi}, J.~S., {Mac Low}, M.-M., {et~al.} 2007, \nat, 448,
  1022

\bibitem[{{Johansen} {et~al.}(2009){Johansen}, {Youdin}, \&
  {Klahr}}]{Johansen_2009}
{Johansen}, A., {Youdin}, A., \& {Klahr}, H. 2009, \apj, 697, 1269

\bibitem[{{Kokubo} \& {Ida}(1996)}]{Kokubo_1996}
{Kokubo}, E., \& {Ida}, S. 1996, \icarus, 123, 180

\bibitem[{{Lambrechts} \& {Johansen}(2012)}]{Lambrechts_2012}
{Lambrechts}, M., \& {Johansen}, A. 2012, \aap, 544, A32

\bibitem[{{Lef{\`e}vre} {et~al.}(2014){Lef{\`e}vre}, {Pagani}, {Juvela},
  {Paladini}, {Lallement}, {Marshall}, {Andersen}, {Bacmann}, {McGehee},
  {Montier}, {Noriega-Crespo}, {Pelkonen}, {Ristorcelli}, \&
  {Steinacker}}]{Lefevre_2014}
{Lef{\`e}vre}, C., {Pagani}, L., {Juvela}, M., {et~al.} 2014, \aap, 572, A20

\bibitem[{{Leger} {et~al.}(1985){Leger}, {Jura}, \& {Omont}}]{Leger_1985}
{Leger}, A., {Jura}, M., \& {Omont}, A. 1985, \aap, 144, 147

\bibitem[{{Leroy} {et~al.}(2013){Leroy}, {Walter}, {Sandstrom}, {Schruba},
  {Munoz-Mateos}, {Bigiel}, {Bolatto}, {Brinks}, {de Blok}, {Meidt}, {Rix},
  {Rosolowsky}, {Schinnerer}, {Schuster}, \& {Usero}}]{Leroy_2013}
{Leroy}, A.~K., {Walter}, F., {Sandstrom}, K., {et~al.} 2013, \aj, 146, 19

\bibitem[{{Levison} {et~al.}(2015){Levison}, {Kretke}, \&
  {Duncan}}]{Levison_2015}
{Levison}, H.~F., {Kretke}, K.~A., \& {Duncan}, M.~J. 2015, \nat, 524, 322

\bibitem[{{Lopez} \& {Fortney}(2014)}]{Lopez_2014}
{Lopez}, E.~D., \& {Fortney}, J.~J. 2014, \apj, 792, 1

\bibitem[{{Lynden-Bell} \& {Pringle}(1974)}]{Lynden-Bell_1974}
{Lynden-Bell}, D., \& {Pringle}, J.~E. 1974, \mnras, 168, 603

\bibitem[{{Mamajek}(2009)}]{Mamajek_2009}
{Mamajek}, E.~E. 2009, in American Institute of Physics Conference Series, Vol.
  1158, American Institute of Physics Conference Series, ed. T.~{Usuda},
  M.~{Tamura}, \& M.~{Ishii}, 3--10

\bibitem[{{Meheut} {et~al.}(2012){Meheut}, {Meliani}, {Varniere}, \&
  {Benz}}]{Meheut_2012}
{Meheut}, H., {Meliani}, Z., {Varniere}, P., \& {Benz}, W. 2012, \aap, 545,
  A134

\bibitem[{{Oishi} {et~al.}(2007){Oishi}, {Mac Low}, \& {Menou}}]{Oishi_2007}
{Oishi}, J.~S., {Mac Low}, M.-M., \& {Menou}, K. 2007, \apj, 670, 805

\bibitem[{{Ormel} {et~al.}(2008){Ormel}, {Cuzzi}, \& {Tielens}}]{Ormel_2008}
{Ormel}, C.~W., {Cuzzi}, J.~N., \& {Tielens}, A.~G.~G.~M. 2008, \apj, 679, 1588

\bibitem[{{Ormel} \& {Klahr}(2010)}]{Ormel_2010}
{Ormel}, C.~W., \& {Klahr}, H.~H. 2010, \aap, 520, A43

\bibitem[{{Raymond} {et~al.}(2004){Raymond}, {Quinn}, \&
  {Lunine}}]{Raymond_2004}
{Raymond}, S.~N., {Quinn}, T., \& {Lunine}, J.~I. 2004, \icarus, 168, 1

\bibitem[{{Raymond} {et~al.}(2006){Raymond}, {Quinn}, \&
  {Lunine}}]{Raymond_2006}
---. 2006, \icarus, 183, 265

\bibitem[{{Ros} \& {Johansen}(2013)}]{Ros_2013}
{Ros}, K., \& {Johansen}, A. 2013, \aap, 552, A137

\bibitem[{{Safronov}(1969)}]{Safronov_1969}
{Safronov}, V.~S. 1969, {Evoliutsiia doplanetnogo oblaka.}

\bibitem[{{Sekiya}(1998)}]{Sekiya_1998}
{Sekiya}, M. 1998, \icarus, 133, 298

\bibitem[{{Shakura} \& {Sunyaev}(1976)}]{Shakura_1976}
{Shakura}, N.~I., \& {Sunyaev}, R.~A. 1976, \mnras, 175, 613

\bibitem[{{Simon} {et~al.}(2012){Simon}, {Beckwith}, \&
  {Armitage}}]{Simon_2012}
{Simon}, J.~B., {Beckwith}, K., \& {Armitage}, P.~J. 2012, \mnras, 422, 2685

\bibitem[{{Simon} {et~al.}(2015){Simon}, {Lesur}, {Kunz}, \&
  {Armitage}}]{Simon_2015}
{Simon}, J.~B., {Lesur}, G., {Kunz}, M.~W., \& {Armitage}, P.~J. 2015, \mnras,
  454, 1117

\bibitem[{{Sirono}(2011{\natexlab{a}})}]{Sirono_2011b}
{Sirono}, S.-i. 2011{\natexlab{a}}, \apjl, 733, L41

\bibitem[{{Sirono}(2011{\natexlab{b}})}]{Sirono_2011a}
---. 2011{\natexlab{b}}, \apj, 735, 131

\bibitem[{{Stoll} \& {Kley}(2014)}]{Stoll_2014}
{Stoll}, M.~H.~R., \& {Kley}, W. 2014, \aap, 572, A77

\bibitem[{{Teague} {et~al.}(2016){Teague}, {Guilloteau}, {Semenov}, {Henning},
  {Dutrey}, {Pi{\'e}tu}, {Birnstiel}, {Chapillon}, {Hollenbach}, \&
  {Gorti}}]{Teague_2016}
{Teague}, R., {Guilloteau}, S., {Semenov}, D., {et~al.} 2016, \aap, 592, A49

\bibitem[{{Turner} {et~al.}(2014){Turner}, {Fromang}, {Gammie}, {Klahr},
  {Lesur}, {Wardle}, \& {Bai}}]{Turner_2014}
{Turner}, N.~J., {Fromang}, S., {Gammie}, C., {et~al.} 2014, Protostars and
  Planets VI, 411

\bibitem[{{van der Marel} {et~al.}(2013){van der Marel}, {van Dishoeck},
  {Bruderer}, {Birnstiel}, {Pinilla}, {Dullemond}, {van Kempen}, {Schmalzl},
  {Brown}, {Herczeg}, {Mathews}, \& {Geers}}]{van_der_Marel_2013}
{van der Marel}, N., {van Dishoeck}, E.~F., {Bruderer}, S., {et~al.} 2013,
  Science, 340, 1199

\bibitem[{{Wada} {et~al.}(2009){Wada}, {Tanaka}, {Suyama}, {Kimura}, \&
  {Yamamoto}}]{Wada_2009}
{Wada}, K., {Tanaka}, H., {Suyama}, T., {Kimura}, H., \& {Yamamoto}, T. 2009,
  \apj, 702, 1490

\bibitem[{{Weidenschilling}(1977)}]{Weidenschilling_1977}
{Weidenschilling}, S.~J. 1977, \mnras, 180, 57

\bibitem[{{Weidenschilling}(1984)}]{Weidenschilling_1984}
---. 1984, \icarus, 60, 553

\bibitem[{{Weingartner} \& {Draine}(2001)}]{Weingartner_2001}
{Weingartner}, J.~C., \& {Draine}, B.~T. 2001, \apj, 548, 296

\bibitem[{{Williams} \& {Cieza}(2011)}]{Williams_2011}
{Williams}, J.~P., \& {Cieza}, L.~A. 2011, \araa, 49, 67

\bibitem[{{Wyatt}(2008)}]{Wyatt_2008}
{Wyatt}, M.~C. 2008, \araa, 46, 339

\bibitem[{{Yang} {et~al.}(2016){Yang}, {Johansen}, \& {Carrera}}]{Yang_2016}
{Yang}, C.-C., {Johansen}, A., \& {Carrera}, D. 2016, ArXiv e-prints,
  arXiv:1611.07014

\bibitem[{{Yang} {et~al.}(2012){Yang}, {Herczeg}, {Linsky}, {Brown},
  {Johns-Krull}, {Ingleby}, {Calvet}, {Bergin}, \& {Valenti}}]{Yang_2012}
{Yang}, H., {Herczeg}, G.~J., {Linsky}, J.~L., {et~al.} 2012, \apj, 744, 121

\bibitem[{{Youdin} \& {Goodman}(2005)}]{Youdin_2005}
{Youdin}, A.~N., \& {Goodman}, J. 2005, \apj, 620, 459

\bibitem[{{Youdin} \& {Shu}(2002)}]{Youdin_2002}
{Youdin}, A.~N., \& {Shu}, F.~H. 2002, \apj, 580, 494

\bibitem[{{Zsom} {et~al.}(2010){Zsom}, {Ormel}, {G{\"u}ttler}, {Blum}, \&
  {Dullemond}}]{Zsom_2010}
{Zsom}, A., {Ormel}, C.~W., {G{\"u}ttler}, C., {Blum}, J., \& {Dullemond},
  C.~P. 2010, \aap, 513, A57

\end{thebibliography}

%%%%%%%%%%%%%%%%%%%%%%%%%%%%%%%%%%%%%%%%%%%%%%%%%%
\end{document}